\documentclass[12pt]{article}
\usepackage{epsf}
\usepackage{color}
\usepackage{amsmath,amssymb,color,graphics,amscd,amsfonts}

\begin{document}

\begin{flushright} 
\end{flushright} 

\vspace{0.1cm}

\begin{Large}
\vspace{1cm}
\begin{center}
{\bf Non-holomorphic multi-matrix gauge invariant operators 
based on Brauer algebra} \\ 
\end{center}
\end{Large}

\vspace{1.4cm}

\begin{center}
{\large {\bf Yusuke Kimura}}   \\ 

\vspace{0.7cm} 
{\sf y.kimura@qmul.ac.uk}
\vspace{0.8cm} 

Centre for Research in String Theory, 
Department of Physics, \\
Queen Mary University of London, \\
Mile End Road, 
London E1 4NS UK \\

\end{center}

\vspace{1cm}

\begin{abstract}
\noindent 
\end{abstract}

We present an orthogonal basis of  
gauge invariant operators 
constructed from 
some complex matrices for the free matrix field, 
where 
operators are expressed with the help of Brauer 
algebra. 
This is a generalisation of our previous work for a signle 
complex matrix. 
We also discuss
the matrix quantum mechanics relevant to ${\cal N}=4$ SYM 
on $S^{3}\times R$.  
A commuting set of conserved operators 
whose eigenstates are given by the orthogonal basis 
is shown by using enhanced symmetries at zero coupling.

\newpage

\section{Introduction}

According to the AdS/CFT correspondence, 
gauge invariant operators correspond 
to physical states in string theory 
and, moreover, the spectrum of states 
is encoded in the anomalous dimension of 
gauge invariant operators. 
Of primary interest is to establish this correspondence 
exactly and to know how the space-time physics 
can be read from correlators of gauge invariant operators. 

The half-BPS states have attracted much attention 
because of some observations. 
The chiral primary sector 
is described by the complex matrix model which can be 
obtained by the dimensional reduction from the original 
${\cal N}=4$ SYM, which leads to the free fermion description of 
this system \cite{0008016,0111222,0403110,0409174,0505129,0507070}.  
The chiral primary operators 
describe KK gravitons and giant gravitons 
\cite{0008016,0003075,0008015,0107119}, 
whose two-point functions are diagonal \cite{0111222}. 
One finds that 
a particular linear combination of 
single traces and multi-traces corresponds 
a single giant graviton or a set of giant gravitons, 
whose information is offered by Young tableau \cite{0111222}. 
A Young diagram plays the role of organising the multi-trace 
structure of this sector and completely determines  
the space-time physics. 

A next question along this line is 
how we can expand the chiral primary sector 
to more general classes of 
gauge invariant operators. 
It is interesting to ask how group theory can be utilised 
to organise gauge invariant operators 
to make the connection to string theory more manifest. 
In \cite{0709.2158}, 
the chiral primary operator was 
extended to include the complex conjugation of a complex matrix, 
and an orthogonal complete set was constructed 
using the free field correlator. 
Because replacing $X$ by $X^{\dagger}$ in chiral 
primary operators means to change the sign of 
angular momentum and the coupling to the background field 
in the dual string side, 
this sector would describe physics containing 
giant gravitons and 
anti-giant gravitons. 
If the interactions are taken into account properly, 
one may expect to see the instability 
originated from tachyons appearing between those branes. 

A mathematical new element in \cite{0709.2158} was to 
use Brauer algebra to organise 
the multi-trace structure of gauge invariant operators 
constructed from $X$ and $X^{\dagger}$. This is a generalisation 
of the fact 
that the symmetric group played a role to organise 
the multi-trace structure of the chiral primary sector. 
The Brauer algebra contains 
the contraction in addition to 
the group algebra of the symmetric group, which 
manages gauge invariant operators 
involving $X$ and $X^{\dagger}$ in a trace. 
The complete set constructed in 
\cite{0709.2158} has more group theoretic labels than 
the chiral primary sector, which is reflected in the fact that  
the operators were conjectured to describe 
a system of branes and anti-branes. 
Other studies 
of diagonalising the two-point functions in the free field limit 
have been reported 
in \cite{0711.0176,0801.2061,0804.2764,0805.3025,0806.1911}. 

In this paper, we continue this line to explore 
non-holomorphic gauge invariant operators built from 
some complex matrices. 
We work out the construction of diagonal two-point functions 
at the free level 
of the matrix field theory
and see what kind of group theoretic structure 
show up. 

Here are the outline of this paper. 
The construction of the orthogonal set of 
gauge invariant operators 
is reviewed in section \ref{sec:reviewbrauer} 
for the chiral primary sector and 
the non-holomorphic extension. 
In section \ref{sec:constructionU(3)brauer}, 
we shall give the construction of an orthogonal set for 
the non-holomorphic multi-matrix sector. 
Section \ref{sec:examples} is given to show some 
examples of our operator. 
The simplest sector contains the Konishi operator. 
In section \ref{sec:qmcasimirs}, the multi-matrix 
quantum mechanics is discussed. 
In our previous paper \cite{0807.3696}, 
we have introduced the concept of enhanced symmetries at 
the free level and 
have given some operators commuting with the Hamiltonian, 
which were 
exploited to measure the spectrum of 
orthogonal bases. 
It will be shown how the orthogonal 
gauge invariant operators constructed in this paper 
are characterised by 
conserved charges based on the enhanced symmetries. 
In section \ref{sec:discussion}, we discuss some future problems. 
Some detailed calculations are shown in appendices.


\section{Diagonal basis of gauge invariant operators 
built from $X$ and $X^{\dagger}$ based on Brauer algebra : Review }
\label{sec:reviewbrauer}

In this section, we review the chiral primary sector 
and the Brauer basis for $X$ and $X^{\dagger}$ sector. 

Operators 
corresponding to the highest weight in the half-BPS sector
are given by 
holomorphic gauge invariant operators made from 
a complex matrix $X$. 
The diagonal basis of operators constructed from $n$ 
matrices is labelled by a Young diagram  
$R$ with $n$ boxes \cite{0111222}. 
We shall review the 
construction of this basis 
from a group theoretical point of view.

An $N\times N$ matrix $X$ can be viewed as 
an endomorphism acting on 
an $N$-dimensional vector space $V$, i.e. $X$ : $V \rightarrow V$. 
The tensor product $X^{\otimes n}$ acts on $V^{\otimes n}$. 
It is useful to introduce the symmetric group $S_{n}$
as a tool to organise both single trace and multi-trace.  
We let 
elements $\sigma$ of the symmetric group $S_{n}$ act on 
$V^{\otimes n}$ 
as the permutations of $n$ vector spaces $V$.   
One easily finds that 
any gauge invariant operator constructed from 
$n$ $X$'s can be expressed as 
$tr_{n}(\sigma X^{\otimes n})
=X_{i_{\sigma(1)}}^{i_{1}}\cdots X_{i_{\sigma(n)}}^{i_{n}}$
where 
the trace $tr_{n}$ is taken in $V^{\otimes n}$. 
It is noted that $tr_{n}(h\sigma h^{-1} X^{\otimes n})$
also gives the same gauge invariant operator for any $h \in S_{n}$, 
which means the conjugacy classes of 
$S_{n}$ classify gauge invariant 
operators.

It is convenient to start with 
the following Schur-Weyl duality to introduce the representation basis, 
\begin{eqnarray}
V^{\otimes n}=\bigoplus_{R}
V_{R}^{U(N)}\otimes V_{R}^{S_{n}}.  
\end{eqnarray}
This comes from the fact that 
the symmetric group $S_{n}$ is the centraliser of 
$U(N)$ on $V^{\otimes n}$.  
$R$ runs over all irreducible representations with 
$n$ boxes satisfying $c_{1}(R)\le N$, 
where $c_{1}(R)$ is the length of the first column of $R$. 
The projection operator 
associated with an irreducible representation $R$ 
can be expressed as an element 
in the group algebra of 
the symmetric group $S_{n}$ as
\begin{eqnarray}
p_{R}=\frac{d_{R}}{n!}\sum_{\sigma\in S_{n}}\chi_{R}(\sigma)\sigma.
\end{eqnarray}
Consider the following 
particular linear combination of $tr_{n}(\sigma X^{\otimes n})$
\begin{eqnarray}
O_{R}(X):=
tr_{n}(p_{R} X^{\otimes n}).
\label{halfbpsbasis}
\end{eqnarray}  
It was shown in \cite{0111222} that it has the 
diagonal two-point function
\footnote{
The two-point function was reproduced by 
$c=1$ tachyon scattering amplitudes in 
\cite{tai}.
}: 
\begin{eqnarray}
\langle O_{R}(X)^{\dagger}O_{S}(X)\rangle 
= 
\delta _{RS}n_{R}!d_{R}DimR ,
\end{eqnarray}
where $d_{R}$ and $DimR$ are the dimension of 
$S_{n}$ and that of $U(N)$ 
associated with the irreducible reprentation $R$. 
This two-point function 
can be shown using 
the propagator of $X_{ij}$:\footnote{
In this paper, we study the theory with gauge group $U(N)$. 
See \cite{0410236,0703202} for the extension to gauge group $SU(N)$. 
}
\begin{eqnarray}
\langle X^{\dagger}_{ij} X_{kl}\rangle
=\delta_{jk}\delta_{il}. 
\end{eqnarray}
The space-time dependence has been ignored because 
it is easily recovered from the conformal invariance, and 
we are mainly interested in the colour and flavour dependence. 
Some properties of the projector are very helpful to calculate
correlation functions \cite{0205221}.

\vspace{0.4cm}

We extend this sector to include $X^{\dagger}$. 
The construction of an orthogonal set 
for non-holomorphic operators 
in the free matrix field theory 
was completed in \cite{0709.2158}
by introducing the Brauer algebra. 
The Brauer algebra is an algebra which has the group algebra of 
$S_{m}\times S_{n}$ and contractions
\footnote{
The contraction $C$ is a linear map from $V\otimes \bar{V}$ 
to itself. 
The action is 
$C v_{i}\otimes \bar{v}_{j}=\delta_{ij}
v_{k}\otimes \bar{v}_{k}$. 
}. 
It is reviewed in section 3 of \cite{0709.2158}. 

We start with the fact that 
any multi-trace gauge invariant operator constructed from $m$ copies of $X$ 
and 
$n$ copies of $X^{\dagger}$ 
can be indicated through 
an element of 
the Brauer algebra as 
$tr_{m,n}(b X^{\otimes m}\otimes X^{\ast \otimes n})$, 
where $tr_{m,n}$ is taken in the space of 
$V^{\otimes m}\otimes \bar{V}^{\otimes n}$. 
At $m=n=1$, 
$b$ takes 
$1$ (the unit element) and $C$ (contraction), which give 
$trXtrX^{\dagger}$ and 
$tr(XX^{\dagger})$, respectively. 
We note that 
$tr_{m,n}(h^{-1}bh X^{\otimes m}\otimes X^{\ast \otimes n})$ 
for any $h\in S_{m}\times S_{n}$ 
provides the same gauge invariant operator
\footnote{
Using the cyclicity property of the trace, 
the conjugate action of $h$ on $b$ 
results in 
the conjugate action on 
$X^{\otimes m}\otimes X^{\ast \otimes n}$, 
which is re-ordering of 
$X$'s and/or that of $X^{\ast}$'s.  }.
This means multi-trace operators are classified 
by the equivalence classes under the conjugation of 
the symmetric group $S_{m}\times S_{n}$.

The Brauer algebra
\footnote{The Brauer algebra is sensitive to $N$. 
For example, we have $C^{2}=NC$ for a contraction $C$. 
} can be introduced as the centraliser of 
$U(N)$ acting on $V^{\otimes m}\otimes \bar{V}^{\otimes n}$, 
\begin{eqnarray}
V^{\otimes m}\otimes \bar{V}^{\otimes n}
=\bigoplus_{\gamma}
V_{\gamma}^{U(N)}\otimes V_{\gamma}^{B_{N}(m,n)}. 
\label{swbrauer} 
\end{eqnarray}
The sum is over irreducible representations 
$\gamma$ of $U(N)$ and $B_{N}(m,n)$. 
This equation follows from the fact that 
Brauer elements commute with the action of $U(N)$ on 
$V^{\otimes m}\otimes \bar{V}^{\otimes n}$. 
The irreducible representation 
$\gamma$ is determined by a set $(\gamma_{+},\gamma_{-},k)$, 
where $\gamma_{+}$ is a partition of $m-k$, 
$\gamma_{-}$ is a partition of $n-k$ 
and $k$ is an integer with $0\le k \le$ min$(m,n)$. 
These definitions provide a constraint 
$c_{1}(\gamma_{+})+c_{1}(\gamma_{-}) \le N$. 
Because the Wick contractions are symbolised in terms of 
the symmetric group $S_{m}\times S_{n}$ 
(see, e.g. (\ref{wickcovariantoperator})), 
it will be convenient to decompose the Brauer algebra 
into the group algebra of $S_{m}\times S_{n}$ 
(which we shall denoted by 
$\mathbb C(S_{m}\times S_{n})$) as 
\begin{eqnarray}
V_{\gamma}^{B_{N}(m,n)}=
\bigoplus_{A}V_{A}^{\mathbb C(S_{m}\times S_{n})}
\otimes V_{\gamma\rightarrow A}.
\label{decomposebrauersym}
\end{eqnarray}
The sum is taken over irreducible representations $A$ of 
the symmetric group, and 
$V_{\gamma\rightarrow A}$ 
represents the space of the multiplicity associated with the decomposition. 
We shall express 
an irreducible representation 
$A$ of $S_{m}\times S_{n}$ 
as a set of a partition of $m$ 
and a partition of $n$ : $(\alpha,\beta)$. 
The multiplicity of the irreducible representation 
$ A=( \alpha , \beta ) $ of  $ \mathbb  C  [ S_m \times S_n ] $
appearing in the irreducible representation
$ \gamma $ of $B_{N}(m,n)$ 
is read from the formula
\begin{eqnarray}
M^{\gamma}_{A}:=
Dim(V_{\gamma\rightarrow A})
=  \sum_{\delta \vdash k } g ( \delta , \gamma_+ ; \alpha ) 
                          g ( \delta , \gamma_- ; \beta ) . 
\label{formulaMultiplicity}
\end{eqnarray}
Here 
$\delta \vdash k$ is shorthand to express that 
$ \delta $ is a partition of $k $. 
The Littlewood-Richardson coefficient 
$g ( \delta , \gamma_+ ; \alpha )$ 
represents 
the multiplicity of the representation 
$\alpha$ appearing in the tensor product of 
the representations $\delta$ and $\gamma_{+}$.
This formula states 
$\gamma_{+}=\alpha$ and 
$\gamma_{-}=\beta$ at $k=0$. 

\vspace{0.2cm}

An orthogonal and complete set introduced in \cite{0709.2158} 
is 
\begin{eqnarray}
O_{A,ij}^{\gamma}(X,X^{\ast})=
:tr_{m,n}(Q_{A,ij}^{\gamma}
X^{\otimes m}\otimes X^{\ast \otimes n}):,
\label{completesetbrauer}
\end{eqnarray}
where $:$ means that 
divergences 
(self-contractions)
associated with defining a composite 
operator 
are omitted, i.e. 
$\langle :tr(\hspace{0.4cm}): \rangle =0$. 
The indices $i,j$ 
run over $1$ to $M_{A}^{\gamma}$, 
behaving like matrix indices as  
$Q_{A,ij}^{\gamma}
Q_{A^{\prime},kl}^{\gamma^{\prime}}
=\delta_{\gamma\gamma^{\prime}}\delta_{AA^{\prime}}
\delta_{jk}
Q_{A,il}^{\gamma}$. 
When the multiplicity is trivial, this operator becomes 
a projector, which happens at $k=0$, $k=m=n$ \cite{YKSRDT}, and so on.

We introduce the restricted character $\chi_{A,ij}^{\gamma}(b)$ 
which is defined by 
\begin{eqnarray}
\chi_{A,ij}^{\gamma}(b)
=\sum_{m_{A}}
\langle 
\gamma \rightarrow A,m_{A},i|b|
\gamma \rightarrow A,m_{A},j
\rangle, 
\label{restrictedcharacter}
\end{eqnarray}
where 
$|\gamma \rightarrow A,m_{A},j\rangle$ is 
a state of the Brauer algebra in the irreducible representation 
$\gamma$ 
associated with the decomposition to 
the subalgebra 
$\mathbb C(S_{m}\times S_{n})$. 
$m_{A}$ represents components in $A$. 
$\chi_{\gamma}(b)=\sum_{A,i}\chi_{A,ii}^{\gamma}(b)$ 
is the character of the Brauer algebra. 
The restricted character enables us to express $Q_{A,ij}^{\gamma}$ 
in terms of 
elements in the Brauer algebra as 
\footnote{
A typo in (122) of \cite{0807.3696} 
is corrected here.}
\begin{eqnarray}
Q_{A,ij}^{\gamma}=t_{\gamma}\sum_{b}\chi_{A,ji}^{\gamma}(b)b^{\ast}.
\label{Qoperator}
\end{eqnarray}
$t_{\gamma}$ is the dimension of the $U(N)$ irreducible representation 
$\gamma$. 
$b^{\ast}$ is the dual element of $b$ which is specified by 
$tr_{m,n}(bb^{\ast})=1$. For more information about this dual element, 
see \cite{0709.2158,ramthesis}. 
The operator $Q_{A,ij}^{\gamma}$ 
commutes with any element of $S_{m}\times S_{n}$:
\begin{eqnarray}
h Q_{A,ij}^{\gamma}  =Q_{A,ij}^{\gamma}h,
\quad h \in \mathbb C (S_{m}\times S_{n}), 
\label{Qcommutesymmetric}
\end{eqnarray}
which can be shown by exploiting $hb^{\ast}h^{-1}=(hbh^{-1})^{\ast}$. 
The property (\ref{Qcommutesymmetric})
is significant to show the diagonal two-point function. 
A projector associated with an irreducible representation 
$\gamma$ can be written down as
\begin{eqnarray}
P^{\gamma}=t_{\gamma}\sum_{b}\chi_{\gamma}(b)b^{\ast}
=\sum_{A,i}Q_{A,ii}^{\gamma}. 
\label{brauercentralprojector}
\end{eqnarray}

\vspace{0.4cm}

Some remarks are in order. 
The free two-point functions are shown to be 
diagonal:
\begin{eqnarray}
\langle 
O_{A,ij}^{\gamma}(X,X^{\ast})^{\dagger}
O_{A^{\prime},i^{\prime}j^{\prime}}^{\gamma^{\prime}}
(X,X^{\ast})
\rangle
 = m!n!d_{A}t_{\gamma} 
\delta _{\gamma \gamma^{\prime}}
\delta _{AA^{\prime}}
\delta _{ii^{\prime}}
\delta _{jj^{\prime}}, 
\end{eqnarray}
where 
$d_{A}$ is the dimension of $S_{m}\times S_{n}$ 
associated with 
the irreducible representation $A$. 


There is a special sub-class of the operator $Q^{\gamma}_{A,ij}$, 
which is parametrised by $k=0$. 
In this case, $\gamma=A$, i.e. 
$\gamma_{+}=\alpha$, $\gamma_{-}=\beta$ 
(see the comment below (\ref{formulaMultiplicity})). 
It is therefore labelled by 
two Young diagrams. 
We shall denote it by 
$P_{\alpha\beta}$. 
Because 
the leading term of $P_{\alpha\beta}$ is  found to be
$p_{\alpha}p_{\beta}$, this gives the product of 
the holomorphic operator and the anti-holomorphic operator 
$O_{\alpha}(X)O_{\beta}(X^{\dagger})$. 
The $k=0$ projector $P_{\alpha\beta}$ also plays an important role 
in the context of 
the large $N$ expansion of two-dimensional Yang-Mills \cite{9301068}. 
Using some properties of the Brauer algebra, 
a new expression of the $SU(N)$ dimension was given, 
leading to a new formulation of two-dimensional 
Yang-Mills \cite{0802.3662}. 

This class of gauge invariant operators does not require the normal 
ordering prescription to make it well-defined 
as composite operators. 
In other words we can show 
\begin{eqnarray}
:tr_{m,n}(P_{\alpha\beta}X^{\otimes m}\otimes X^{\ast \otimes n}):
=tr_{m,n}(P_{\alpha\beta}X^{\otimes m}\otimes X^{\ast \otimes n}). 
\end{eqnarray}
Considering the fact that 
the Wick contraction between 
an $X$ and an $X^{\ast}$ is performed by a contraction $C$, 
$C P_{\alpha\beta}=0$ is a sufficient condition for the above 
equation. 
It indeed follows from the fact
that the projector $P_{\alpha\beta}$ is orthogonal to 
the other projectors relevant to $k\neq0$. 

We conclude this section by showing 
the simplest example of $m=n=1$:
\begin{eqnarray}
&&
trXtrX^{\dagger}-\frac{1}{N}tr(XX^{\dagger}), 
\quad
\frac{1}{N}tr(XX^{\dagger}).
\end{eqnarray}
The first one is labelled by $k=0$ and $A=([1],[1])$, 
while the second one is by $\gamma=(\emptyset,\emptyset,1)$ 
and $A=([1],[1])$.

\section{Diagonal basis of non-holomorphic multi-matrix 
gauge invariant operators}
\label{sec:constructionU(3)brauer}
In the previous section, we have reviewed 
a specific set of gauge invariant operators 
made from $X$ and $X^{\dagger}$ that are engineered by 
the Brauer algebra. 
In this section we generalise this to multi-matrix models. 
We shall present an orthogonal set of 
operators composed of $X_{a}$ and $X_{a}^{\dagger}$, 
where $a=1,\cdots,p$, 
at the free coupling. 
This has an additional flavour index compared 
to the previous case. 
Hence we first work out the flavour structure, 
and later we will move to the colour structure. 
This procedure 
exploits the approach of 
\cite{0711.0176} to deal with global indices. 
When $p$ is $3$, 
it is relevant to the 
$SO(6)$ sector of 
${\cal N}=4$ four-dimensional super Yang-Mills 
theory.

\subsection{Gauge covariant operators in representation basis}

We start with the following gauge covariant operator 
\begin{eqnarray}
(O_{\vec{a},\vec{b}})^{I}_{J}:= 
(X_{a_{1}})^{i_{1}}_{j_{1}}
\otimes \cdots \otimes 
(X_{a_{m}})^{i_{m}}_{j_{m}}
\otimes 
(X_{b_{1}}^{\ast})^{i_{m+1}}_{j_{m+1}}
\otimes \cdots \otimes 
(X_{b_{n}}^{\ast})^{i_{m+n}}_{j_{m+n}},
\label{NaiveBasisCovariantOperator}
\end{eqnarray}
where $a_{i},b_{i}=1,2,\ldots,p$. 
We will rewrite the flavour structure 
using a representation basis. 

Let $V_{F}$ be the space of the fundamental representation 
of $U(p)$. 
The following Schur-Weyl duality is relevant to the flavour structure: 
\begin{eqnarray}
V_{F}^{\otimes m}\otimes \bar{V}_{F}^{\otimes n}
=\bigoplus_{\Lambda}V_{\Lambda}^{U(p)}\otimes V_{\Lambda}^{B_{p}(m,n)}.
\label{swu3}
\end{eqnarray}
Here the sum runs over
irreducible representations of 
$U(p)$ and $B_{p}(m,n)$. 
The representation $\Lambda$ is labelled by 
a set $(\Lambda_{+},\Lambda_{-},l)$, where 
$l$ is an integer with $0\le l \le$ min$(m,n)$, and 
$\Lambda_{+}$ and $\Lambda_{-}$ 
are given by a partition of $m-l$ 
and a partition of $n-l$.
It is noted that 
this Brauer algebra $B_{p}(m,n)$ should not be confused with 
another Brauer algebra $B_{N}(m,n)$ which is
relevant for the colour structure.  
 
The group algebra of $S_{m}\times S_{n}$, $\mathbb C(S_{m}\times S_{n})$, 
is a subalgebra of $B_{p}(m,n)$, hence 
we consider the decomposition of $B_{p}(m,n)$ into 
$\mathbb C(S_{m}\times S_{n})$:
\begin{eqnarray}
V_{\Lambda}^{B_{p}(m,n)}
=\bigoplus_{\Lambda_{1}}
V_{\Lambda_{1}}^{\mathbb C(S_{m}\times S_{n})}\otimes
V_{\Lambda \rightarrow \Lambda_{1}},
\label{SmSninB}
\end{eqnarray}
where 
$\Lambda_{1}$ runs over irreducible representations of 
$\mathbb C(S_{m}\times S_{n})$. The second factor in the right-hand side 
represents the space of the 
multiplicity arising from this decomposition. 
Combining (\ref{swu3}) and (\ref{SmSninB}), 
we have the following equation for the flavour structure 
\begin{eqnarray}
V_{F}^{\otimes m}\otimes \bar{V}_{F}^{\otimes n}
=\bigoplus_{\Lambda,\Lambda_{1}}V_{\Lambda}^{U(p)}\otimes 
V_{\Lambda_{1}}^{\mathbb C(S_{m}\times S_{n})}
\otimes 
V_{\Lambda\rightarrow \Lambda_{1}}.
\label{scdecomposed}
\end{eqnarray}
Based on this decomposition, 
one may introduce 
a covariant operator 
in a representation basis  
as 
\begin{eqnarray}
O_{\vec{a},\vec{b}}
&=&
\sum_{\Lambda,M_{\Lambda},\Lambda_{1},m_{\Lambda_{1}},\tau}
C_{\vec{a},\vec{b}}^{\Lambda,M_{\Lambda},\Lambda_{1},m_{\Lambda_{1}},\tau}
O_{\Lambda,M_{\Lambda},\Lambda_{1},m_{\Lambda_{1}},\tau}.
\end{eqnarray}
$M_{\Lambda}$ represents 
states in the irreducible representation $\Lambda$ of 
$U(p)$, and 
$m_{\Lambda_{1}}$ runs over states in the $\Lambda_{1}$. 
$\tau$ is an index running over the multiplicity of 
$\Lambda_{1}$ in $\Lambda$. 
The inverse is 
\begin{eqnarray}
O_{\Lambda,M_{\Lambda},\Lambda_{1},m_{\Lambda_{1}},\tau}
=\sum_{\vec{a},\vec{b}}C_{\Lambda,M_{\Lambda},
\Lambda_{1},m_{\Lambda_{1}},\tau}^{\vec{a},\vec{b}}
O_{\vec{a},\vec{b}}. 
\label{operatorrepbasis}
\end{eqnarray}

We now calculate the free two-point function of the 
operator. 
Using 
\begin{eqnarray}
\langle (X^{\dagger}_{a})_{ij} (X_{b})_{kl}\rangle
=\delta_{ab}\delta_{jk}\delta_{il}, 
\end{eqnarray}
we get for $O_{\vec{a},\vec{b}}$
\begin{eqnarray}
\langle : (O_{\vec{a},\vec{b}}^{\dagger})_{J}^{I}:
: (O_{\vec{a^{\prime}},\vec{b^{\prime}}})_{L}^{K}:
\rangle
=\sum_{\sigma \in S_{m}\times S_{n}}
\prod_{k=1}^{m}\delta_{a_{k}a^{\prime}_{\sigma(k)}} 
\prod_{l=1}^{n}\delta_{b_{l}b^{\prime}_{\sigma(l)}} 
(\sigma)_{J}^{K}(\sigma^{-1})_{L}^{I},
\label{wickcovariantoperator}
\end{eqnarray}
where 
\begin{eqnarray}
(\sigma)_{J}^{K}=
(\sigma)_{j_{1}\cdots j_{m+n}}^{k_{1}\cdots k_{m+n}}
:=\delta^{k_{1}}_{j_{\sigma(1)}}
\cdots 
\delta^{k_{m+n}}_{j_{\sigma(m+n)}}.
\end{eqnarray}
 
The two-point function of the representation basis 
(\ref{operatorrepbasis}) can be computed as 
\begin{eqnarray}
&&\langle 
: (O_{\Lambda,M_{\Lambda},\Lambda_{1},m_{\Lambda_{1}},
\tau}^{\dagger})_{J}^{I}:
: (O_{\Lambda^{\prime},M_{\Lambda^{\prime}}^{\prime},\Lambda_{1}^{\prime},
m_{\Lambda_{1}^{\prime}}^{\prime},\tau^{\prime}})_{L}^{K}:
\rangle \cr
&=&
\delta_{\Lambda \Lambda^{\prime}}
\delta_{M_{\Lambda}M_{\Lambda^{\prime}}^{\prime}}
\delta_{\Lambda_{1}\Lambda_{1}^{\prime}}
\delta_{\tau\tau^{\prime}}
\sum_{\sigma \in S_{m}\times S_{n}}
D^{\Lambda_{1}}_{m_{\Lambda_{1}}m_{\Lambda_{1}^{\prime}}^{\prime}}(\sigma)
(\sigma)_{J}^{K}(\sigma^{-1})_{L}^{I}.
\label{covariant2pt}
\end{eqnarray}
This will be proved in appendix \ref{proofdiagonaltwopointcovariant}. 

\subsection{Gauge invariant operators in representation basis}

All gauge invariant operators constructed from the 
covariant operators 
are expressed by 
$tr_{m,n}(bO_{\Lambda,M_{\Lambda},\Lambda_{1},m_{\Lambda_{1}},\tau})$. 
To rewrite the colour structure in terms of a representation basis, 
we again use the decomposition given in 
(\ref{swbrauer}) and (\ref{decomposebrauersym}):
\begin{eqnarray}
V^{\otimes m}\otimes \bar{V}^{\otimes n}
=\bigoplus_{\gamma,A}
V_{\gamma}^{U(N)}\otimes 
V_{A}^{\mathbb C(S_{m}\times S_{n})}
\otimes V_{\gamma\rightarrow A}.
\end{eqnarray}
We thus have two kinds of 
irreducible representations of the symmetric group. 
One, $\Lambda$, is responsible for the global indices, 
and the other, $A$, is for the colour indices. 
These two kinds of representations come from 
two different actions of the symmetric group which 
are related each other 
as we shall discuss later 
(see (\ref{twosymrelation}) and appendix \ref{sec:symmetryCG}). 
We now consider 
the inner tensor product of $A\otimes A$ decomposing into 
$\Lambda_{1}$ and 
denote the multiplicity of $\Lambda_{1}$ by 
$\tau_{\Lambda_{1}}$. 
Introducing the Clebsch-Gordan coefficient 
\footnote{
See \cite{Hamermesh} for some properties of 
the Clebsch-Gordan coefficient.
} 
$C^{\tau_{\Lambda_{1}},\Lambda_{1},m_{\Lambda_{1}}}
_{A,m_{A},A,m_{A}^{\prime}}$ 
associated with this decomposition, 
we propose the 
following gauge invariant operator
\begin{eqnarray}
{\cal O}_{\Lambda,M_{\Lambda},\Lambda_{1},\tau}
^{\gamma,A,ij,\tau_{\Lambda_{1}}}
=
\sum_{m_{A},m_{A}^{\prime},m_{\Lambda_{1}}}
C^{\tau_{\Lambda_{1}},\Lambda_{1},m_{\Lambda_{1}}}
_{A,m_{A},A,m_{A}^{\prime}}
\sum_{b\in B_{N}(m,n)}
D^{\gamma}_{A,m_{A},j,m_{A}^{\prime},i}(b^{\ast})
:
tr_{m,n}(bO_{\Lambda,M_{\Lambda},\Lambda_{1},m_{\Lambda_{1}},\tau})
:.
\label{gaugeinvariant}
\end{eqnarray}
Here
\begin{eqnarray}
D^{\gamma}_{A,m_{A},j,m_{A}^{\prime},i}(b)
&:=& \langle \gamma\rightarrow A,m_{A},j |b|
\gamma\rightarrow A,m_{A}^{\prime},i \rangle . 
\end{eqnarray}
A remark is 
\begin{eqnarray}
\chi_{A,ji}^{\gamma}(b)=
\sum_{m_{A}}
D^{\gamma}_{A,m_{A},j,m_{A},i}(b)
\end{eqnarray}
is the restricted character in (\ref{restrictedcharacter}).

We can show that the two-point function of 
the operator (\ref{gaugeinvariant}) is diagonal:
\begin{eqnarray}
\langle 
{\cal O}^{\gamma,A,ij,\tau_{\Lambda_{1}}}
_{\Lambda,M_{\Lambda},\Lambda_{1},\tau}{}^{\dagger}
{\cal O}^{\gamma^{\prime},A^{\prime},
i^{\prime}j^{\prime},\tau_{\Lambda_{1}^{\prime}}^{\prime}}
_{\Lambda^{\prime},M_{\Lambda^{\prime}}^{\prime},
\Lambda_{1}^{\prime},\tau^{\prime}}
\rangle 
=
m!n!d_{\Lambda_{1}}
\frac{1}{t^{\gamma}}
\delta_{\gamma\gamma^{\prime}}
\delta_{AA^{\prime}}
\delta_{\Lambda \Lambda^{\prime}}
\delta_{M_{\Lambda}M_{\Lambda^{\prime}}^{\prime}}
\delta_{\Lambda_{1}\Lambda_{1}^{\prime}}
\delta_{
\tau_{\Lambda_{1}}
\tau_{\Lambda_{1}^{\prime}}^{\prime}}
\delta_{\tau\tau^{\prime}} 
\delta_{ii^{\prime}}\delta_{jj^{\prime}}
.
\label{diagonaltwopoint}
\end{eqnarray}
This is the main result of this paper. 
We shall give the proof to the appendix 
\ref{proofdiagonaltwopoint}. 

\vspace{0.4cm}

For $p=1$, the operator should coincide 
to the operator seen in the previous section. 
In this case, $\Lambda$
is a one-dimensional representation, so 
the covariant operator is labelled by $\Lambda$ alone 
i.e. the number of $X$ and that of $X^{\ast}$. 
The Clebsh-Gordan for this case is given by
\begin{eqnarray}
C_{A,m_{A},A,m_{A}^{\prime}}
=\frac{1}{\sqrt{d_{A}}}\delta_{m_{A}m_{A}^{\prime}}.
\end{eqnarray}
We thus obtain 
\begin{eqnarray}
{\cal O}
_{\gamma,A,ij}=
\frac{1}{\sqrt{d_{A}}}
\sum_{b}
\chi^{\gamma}_{A,ji}(b^{\ast})
tr_{m,n}(bX^{\otimes m}\otimes X^{\ast \otimes n}),
\end{eqnarray}
which is equivalent to (\ref{completesetbrauer}) 
up to the normalisation factor. 

\vspace{0.4cm}

Before proceeding to the next subsection, we 
summarise the relation between the two 
actions of the symmetric group. 
Two kinds of Brauer algebras have been introduced, 
one is relevant for the flavour structure, and the other 
is for the colour structure. Both Brauer algebras 
contain the same sub-algebra, that is,  
the group algebra of $S_{m}\times S_{n}$, 
but they admit different actions on the operator. 
The permutation acting on the flavour indices 
is equivalent to re-ordering of $X$'s and/or that of $X^{\ast}$'s. 
If we act with permutations on upper colour indices 
and lower colour indices simultaneously, 
we effectively get permutations on flavour indices:
\begin{eqnarray}
\left(O_{\vec{a},\vec{b}}
\right)^{\sigma(I)}_{\sigma(J)}=
\left(O_{\sigma(\vec{a}),\sigma(\vec{b})}
\right)^{I}_{J}. 
\label{twosymrelation}
\end{eqnarray}
In appendix \ref{sec:symmetryCG} 
we shall confirm that 
the gauge invariant operator respects this symmetry. 


\subsection{Operator in the $k=0$ representation of $\gamma$}

Let us now study the special class of 
the operator where 
the representation of $\gamma$ is specified by $k=0$. 
The multiplicity indices related to 
the decomposition $\gamma \rightarrow A$ can go away 
because $\gamma=A$ 
from the formula (\ref{formulaMultiplicity}) at $k=0$. 
So the operator is simplified to be 
\begin{eqnarray}
{\cal O}_{\Lambda,M_{\Lambda},\Lambda_{1},\tau}
^{A,\tau_{\Lambda_{1}}}
&=&
C^{\tau_{\Lambda_{1}},\Lambda_{1},m_{\Lambda_{1}}}_{A,m_{A},A,m_{A}^{\prime}}
\sum_{b\in B_{N}(m,n)}
D^{A}_{m_{A},m_{A}^{\prime}}(b)
tr_{m,n}(b^{\ast} O_{\Lambda,M_{\Lambda},\Lambda_{1},m_{\Lambda_{1}},\tau})
\cr
&=&
C^{\tau_{\Lambda_{1}},\Lambda_{1},m_{\Lambda_{1}}}_{A,m_{A},A,m_{A}^{\prime}}
\sum_{\alpha \in S_{m}\times S_{n}}
D^{A}_{m_{A},m_{A}^{\prime}}(\alpha)
tr_{m,n}(\alpha^{\ast} 
O_{\Lambda,M_{\Lambda},\Lambda_{1},m_{\Lambda_{1}},\tau})
\cr
&=&
C^{\tau_{\Lambda_{1}},\Lambda_{1},m_{\Lambda_{1}}}_{A,m_{A},A,m_{A}^{\prime}}
\sum_{\alpha \in S_{m}\times S_{n}}
D^{A}_{m_{A},m_{A}^{\prime}}(\alpha)
tr_{m,n}(1^{\ast}\alpha^{-1} 
O_{\Lambda,M_{\Lambda},\Lambda_{1},m_{\Lambda_{1}},\tau}).
\label{k=0simplified}
\end{eqnarray}
$D^{A}_{m_{A},m_{A}^{\prime}}(C)=0$ has been used to get the second equality.
The third equality follows from the formula
$\alpha^{\ast}=(1^{\ast})\alpha^{-1}$ 
for $\alpha \in S_{m}\times S_{n}$ which is 
derived in \cite{0709.2158}.  

For the complete set of the 
$X$ and $X^{\dagger}$ sector, 
we could show that 
operators corresponding to the $k=0$ representation 
do not require 
the normal ordering prescription because 
short distance singularities arising from 
self-contractions vanish. 
This property also holds in this case. 
Using the fact that 
the Wick contraction between an $X$ and an $X^{\ast}$ 
can be expressed by a contraction $C$, 
we show 
that 
replacing a set of an $X$ and an $X^{\ast}$ by
a contraction $C$ in 
(\ref{k=0simplified}) vanishes:
\begin{eqnarray}
&&
\sum_{\alpha}
D^{A}_{m_{A},m_{A}^{\prime}}(\alpha)
tr_{m,n}(C1^{\ast}\alpha^{-1} 
O^{\prime})
\cr
&=&
\sum_{\alpha}
D^{A}_{m_{A},m_{A}^{\prime}}(\alpha)
\sum_{\gamma^{\prime}}
\chi^{\gamma^{\prime}}(C1^{\ast}\alpha^{-1} )
tr_{m,n}(P^{\gamma^{\prime}}
O^{\prime})
\cr
&=&
\sum_{\alpha}\sum_{\gamma^{\prime}}
D^{A}_{m_{A},m_{A}^{\prime}}(\alpha)
D^{\gamma^{\prime}}_{IJ}(C1^{\ast})
D^{\gamma^{\prime}}_{JI}(\alpha^{-1})
tr_{m,n}(P^{\gamma^{\prime}}
O^{\prime})
\cr
&\propto&
D^{A}_{m_{A},m_{A}^{\prime}}(C1^{\ast})=0.
\end{eqnarray}
Here 
the first equality comes from the decomposition (\ref{swbrauer}), 
and we have used the orthogonality of the representation matrix 
(\ref{orthogonalrepresentationbrauer}) in the last step. 
In this way, 
we can demonstrate that the normal ordering 
prescription is no longer needed for 
gauge invariant operators belonging to the $k=0$ representation.  

\vspace{0.2cm}

Another property of the $k=0$ sector is 
that the leading term contains 
the product of a holomorphic operator
and an anti-holomorphic operator. 
We shall show that 
the operator constructed in \cite{0711.0176} 
will be reproduced as a subset of our operator. 
To show this, we use an explicit expression of $1^{\ast}$: 
\begin{eqnarray}
1^{\ast}=\frac{1}{N^{m+n}}\Omega_{m}^{-1}\Omega_{n}^{-1}+\cdots,
\end{eqnarray}
where 
$\Omega_{m}$ is a central element
in the group algebra of $S_{m}$, and 
$\cdots$ are terms which are not 
in $\mathbb C(S_{m}\times S_{n})$
but in $B_{N}(m,n)$.  
See appendix $A$ in \cite{0709.2158} 
about more concrete form of $1^{\ast}$ for some examples. 
We substitute 
the leading term of $1^{\ast}$ into 
(\ref{k=0simplified}) and 
restrict to the $l=0$ sector of the representation $\Lambda$. 
This implies we do not see terms 
in which flavour indices are contracted 
between $X$ and $X^{\dagger}$.  
This is the case of
$\Lambda=\Lambda_{1}$, so the index $\tau$ 
can be suppressed. 

The operator 
(\ref{k=0simplified}) with the above restrictions taken into account 
becomes 
\begin{eqnarray}
&&
\frac{1}{N^{m+n}}
C^{\tau_{\Lambda_{1}},\Lambda_{1},m_{\Lambda_{1}}}_{A,m_{A},A,m_{A}^{\prime}}
\sum_{\alpha \in S_{m}\times S_{n}}
D^{A}_{m_{A},m_{A}^{\prime}}(\alpha)
tr_{m,n}(
\Omega_{m}^{-1}\Omega_{n}^{-1}
\alpha^{-1} 
O_{\Lambda_{1},m_{\Lambda_{1}}})
\cr
&=&
\frac{1}{N^{m+n}}
C^{\tau_{\Lambda_{1}},\Lambda_{1},m_{\Lambda_{1}}}_{A,m_{A},A,m_{A}^{\prime}}
\sum_{\alpha \in S_{m}\times S_{n}}
D^{A}_{m_{A},m_{A}^{\prime}}(
\Omega_{m}^{-1}\Omega_{n}^{-1}
\alpha)
tr_{m,n}(
\alpha^{-1} 
O_{\Lambda_{1},m_{\Lambda_{1}}})
\cr
&=&
\frac{1}{N^{m+n}}
C^{\tau_{\Lambda_{1}},\Lambda_{1},m_{\Lambda_{1}}}_{A,m_{A},A,m_{A}^{\prime}}
\frac{1}{d_{A}}\chi_{A}\left(\Omega_{m}^{-1}\Omega_{n}^{-1}\right)
\sum_{\alpha \in S_{m}\times S_{n}}
D^{A}_{m_{A},m_{A}^{\prime}}(\alpha)
tr_{m,n}(
\alpha^{-1} 
O_{\Lambda_{1},m_{\Lambda_{1}}})
\cr
&=&
\frac{1}{N^{m+n}}
C^{\tau_{\Lambda_{1}},\Lambda_{1},m_{\Lambda_{1}}}_{A,m_{A},A,m_{A}^{\prime}}
\frac{1}{d_{A}}\chi_{A}\left(\Omega_{m}^{-1}\Omega_{n}^{-1}\right)
\sum_{\sigma \in S_{m},\tau \in \times S_{n}}
D^{A}_{m_{A},m_{A}^{\prime}}(\sigma\otimes \tau)
tr_{m,n}(
\sigma^{-1}\otimes \tau^{-1} 
O_{\Lambda_{1},m_{\Lambda_{1}}}).
\nonumber
\end{eqnarray}
Each factor can be factorised into the $S_{m}$ part and the
$S_{n}$ part. For example, 
the character can be rewritten as 
\begin{eqnarray}
\chi_{A}\left(\Omega_{m}^{-1}\Omega_{n}^{-1}\right)
&=&
\chi_{R}\left(\Omega_{m}^{-1})\chi_{S}(\Omega_{n}^{-1}\right)
\cr
&=&
\frac{1}{m!n!}\frac{N^{m+n}d_{R}^{2}d_{S}^{2}}{DimRDimS},
\end{eqnarray}
where we have expressed 
$A$ as $(R,S)$ where $R$ is an irreducible representation of 
$S_{m}$ and $S$ is an irreducible representation of $S_{n}$,  
and  
the formula 
$DimR=\frac{N^{m}}{m!}\chi_{R}(\Omega_{m})$ has been used. 
Taking this factorisation into account, 
we find that 
the operator in the $k=0$ representation of $\gamma$ 
and the $l=0$ representation of $\Lambda$ contains the product of  
the holomorphic multi-matrix operator 
and the anti-holomorphic multi-matirx operator 
presented in 
\cite{0711.0176,0806.1911}.

\section{Examples}
\label{sec:examples}
In this section, we present explicit forms of our operator 
for two simple cases. 
We will set $p=3$ to see a connection to the ${\cal N}=4$ SYM. 

\subsection{$m=1$, $n=1$}
In this case, 
there are only two cases for 
$\Lambda=(\Lambda_{+},\Lambda_{-},l)$, i.e. 
$([1],[1],0)$ and $(\emptyset,\emptyset,1)$. 
We call them $l=0$ and $l=1$ because 
the integer $l$ completely identifies them. 

The bases $|\Lambda,M_{\Lambda}\rangle$ are
\begin{eqnarray}
&&
|l=0 \rangle
=
|a,b\rangle -\frac{1}{3}\delta_{ab}\sum_{c=1}^{3}|c,c\rangle, 
\cr
&&
|l=1 \rangle
=
\frac{1}{3}\delta_{ab}
\sum_{c=1}^{3}|c,c\rangle.
\end{eqnarray}
Using the above states, 
the Clebsch-Gordan coefficient 
$C_{\Lambda,M_{\Lambda}}^{a^{\prime},b^{\prime}}
=\langle a^{\prime},b^{\prime}|\Lambda,M_{\Lambda}\rangle$
can be calculated as 
\begin{eqnarray}
&&
C^{a^{\prime},b^{\prime}}_{l=0}=
\langle a^{\prime},b^{\prime}|l=0 
\rangle =\delta_{aa^{\prime}}\delta_{bb^{\prime}}
-\frac{1}{3}\delta_{ab}\delta_{a^{\prime}b^{\prime}},
\cr
&&
C^{a^{\prime},b^{\prime}}_{l=1}=
\langle a^{\prime},b^{\prime}|l=1 
\rangle =\frac{1}{3}\delta_{ab}\delta_{a^{\prime}b^{\prime}}, 
\end{eqnarray}
which yield covariant operators 
$O_{\Lambda,M_{\Lambda}}=
C_{\Lambda,M_{\Lambda}}^{a^{\prime},b^{\prime}}
(X_{a^{\prime}}\otimes X_{b^{\prime}})$:
\begin{eqnarray}
&&
O_{l=0}=
X_{a}\otimes X_{b}^{\ast}
-\frac{1}{3}\delta_{ab}
\sum_{c^{\prime}}X_{c^{\prime}}\otimes X_{c^{\prime}}^{\ast},
\cr
&&
O_{l=1}=
\frac{1}{3}
\delta_{ab}
\sum_{c^{\prime}}X_{c^{\prime}}\otimes X_{c^{\prime}}^{\ast}.
\end{eqnarray}

We next work out the gauge invariant operator. 
For this case 
the operator (\ref{gaugeinvariant}) is simplified to 
\begin{eqnarray}
{\cal O}^{\gamma}_{\Lambda,M_{\Lambda}}
=
\sum_{b\in B_{N}(1,1)}D^{\gamma}(b^{\ast})tr_{1,1}(bO_{\Lambda,M_{\Lambda}})
=
\frac{1}{t_{\gamma}}tr_{1,1}(P^{\gamma}O_{\Lambda,M_{\Lambda}}).
\end{eqnarray}
$\gamma$ is specified by an integer $k$, which takes either 
$k=0$ or $k=1$. 
$P^{\gamma}$ is a projector associated with $\gamma$:
\begin{eqnarray}
 P^{(k=0)}=1-\frac{C}{N}, \quad P^{(k=1)}=\frac{C}{N}.
\end{eqnarray}
Re-normalising the operators as 
${\cal O}^{\gamma}\rightarrow t_{\gamma}{\cal O}^{\gamma}$ 
for convenience, we 
reach the following gauge invariant operators
\begin{eqnarray}
{\cal O}_{k=0,l=0}&=&
(trX_{a})(trX_{b}^{\dagger})-\frac{1}{N}tr(X_{a}X_{b}^{\dagger})
-\frac{1}{3}\delta_{ab}(trX_{c})(trX_{c}^{\dagger})
+\frac{1}{3N}\delta_{ab}tr(X_{c}X_{c}^{\dagger})
\cr
{\cal O}_{k=1,l=0}&=&
\frac{1}{N}\left(
tr(X_{a}X_{b}^{\dagger})
-\frac{1}{3}\delta_{ab}tr(X_{c}X_{c}^{\dagger})
\right)
\cr
{\cal O}_{k=0,l=1}
&=&
\frac{1}{3}\delta_{ab}
\left(
(trX_{c})(trX_{c}^{\dagger})-\frac{1}{N}tr(X_{c}X_{c}^{\dagger})
\right)
\cr
{\cal O}_{k=1,l=1}
&=&
\frac{1}{3N}\delta_{ab}
tr(X_{c}X_{c}^{\dagger}). 
\end{eqnarray}
In fact, these are not eigenstates 
of the one-loop dilatation operator
\footnote{
The one-loop dilatation operator in the so(6) sector will be
rewritten in terms of the complex variables 
in appendix \ref{sec:hamiltoniancomplex}. }. 
Correct eigenstates at one-loop are 
\begin{eqnarray}
&& {\cal O}_{k=0,l=0} \cr
&& {\cal O}_{k=1,l=0} \cr
&& {\cal O}_{k=0,l=1} +{\cal O}_{k=1,l=1}
=\frac{1}{3}
\delta_{ab}
(trX_{c})(trX_{c}^{\dagger})
\cr
&& \frac{N^{2}-1}{N}{\cal O}_{k=1,l=1}-\frac{1}{N}{\cal O}_{k=0,l=1}=
\frac{1}{3}
\delta_{ab}
\left(
tr(X_{c}X_{c}^{\dagger}) -\frac{1}{N}(trX_{c})(trX_{c}^{\dagger})
\right)
\end{eqnarray}
with eigenvalues $0$, $0$, $0$, $6/\sqrt{3}$, respectively.
The first three are 
operators in the short multiplets. 
The last one is the Konishi operator. 
The $l=0$ representations are not mixed up with 
the $l=1$ representations 
because the Hamiltonian commutes with the Brauer algebra 
governing the flavour structure.

\subsection{$m=2$, $n=1$}
We next show the case at $m=2$, $n=1$. 
States $| \Lambda,\Lambda_{1}  \rangle$ are
\begin{eqnarray}
|[2],[1],0 \rangle
&=&
\frac{1}{2}
(|a,b,c\rangle +|b,a,c\rangle )
-\frac{1}{8}\delta_{ac}
\sum_{d=1}^{3}(|d,b,d\rangle+|b,d,d\rangle)
\cr
&&
-\frac{1}{8}\delta_{bc}
\sum_{d=1}^{3}(|a,d,d\rangle+|d,a,d\rangle)
\cr
|[1,1],[1],0 \rangle
&=&
\frac{1}{2}
(|a,b,c\rangle -|b,a,c\rangle )
-\frac{1}{4}\delta_{ac}
\sum_{d=1}^{3}(|d,b,d\rangle-|b,d,d\rangle)
\cr
&&
-\frac{1}{4}\delta_{bc}
\sum_{d=1}^{3}(|a,d,d\rangle-|d,a,d\rangle)
\cr
|([1],\emptyset,1), ([2],[1]) \rangle
&=&
\frac{1}{8}\delta_{ac}
\sum_{d=1}^{3}(|d,b,d\rangle+|b,d,d\rangle)
+\frac{1}{8}\delta_{bc}
\sum_{d=1}^{3}(|a,d,d\rangle+|d,a,d\rangle)
\cr
|([1],\emptyset,1), ([1,1],[1]) \rangle
&=&
\frac{1}{4}\delta_{ac}
\sum_{d=1}^{3}(|d,b,d\rangle-|b,d,d\rangle)
+\frac{1}{4}\delta_{bc}
\sum_{d=1}^{3}(|a,d,d\rangle-|d,a,d\rangle).
\nonumber
\end{eqnarray}
Projectors in 
(\ref{centralprojector21}) and (\ref{symmetricprojector21})
with $N$ replaced by $3(=p)$
are useful to calculate the above equations. 
Recall that 
$\Lambda=\Lambda_{1}$ for the $l=0$ representation. 
Then the Clebsch-Gordan coefficients are 
\begin{eqnarray}
C^{a^{\prime},b^{\prime}}_{[2],[1],l=0}
&=&
\frac{1}{2}
(\delta_{aa^{\prime}}\delta_{bb^{\prime}}\delta_{cc^{\prime}}
+\delta_{ab^{\prime}}\delta_{ba^{\prime}}\delta_{cc^{\prime}})
-\frac{1}{8}\delta_{ac}
(\delta_{a^{\prime}c^{\prime}}\delta_{bb^{\prime}}
+\delta_{b^{\prime}c^{\prime}}\delta_{ba^{\prime}})
\cr
&&
-\frac{1}{8}\delta_{bc}
(\delta_{b^{\prime}c^{\prime}}\delta_{aa^{\prime}}
+\delta_{a^{\prime}c^{\prime}}\delta_{ab^{\prime}})
\cr
C^{a^{\prime},b^{\prime}}_{[1,1],[1],l=0}
&=&
\frac{1}{2}
(\delta_{aa^{\prime}}\delta_{bb^{\prime}}\delta_{cc^{\prime}}
-\delta_{ab^{\prime}}\delta_{ba^{\prime}}\delta_{cc^{\prime}})
-\frac{1}{4}\delta_{ac}
(\delta_{a^{\prime}c^{\prime}}\delta_{bb^{\prime}}
-\delta_{b^{\prime}c^{\prime}}\delta_{ba^{\prime}})
\cr
&&
-\frac{1}{4}\delta_{bc}
(\delta_{b^{\prime}c^{\prime}}\delta_{aa^{\prime}}
-\delta_{a^{\prime}c^{\prime}}\delta_{ab^{\prime}})
\cr
C^{a^{\prime},b^{\prime}}_{l=1,\Lambda_{1}=([2],[1])}
&=&
\frac{1}{8}\delta_{ac}
(\delta_{bb^{\prime}}\delta_{a^{\prime}c^{\prime}}
+\delta_{ba^{\prime}}\delta_{b^{\prime}c^{\prime}})
+\frac{1}{8}\delta_{bc}
(
\delta_{aa^{\prime}}\delta_{b^{\prime}c^{\prime}}
+\delta_{ab^{\prime}}\delta_{a^{\prime}c^{\prime}}
)
\cr
C^{a^{\prime},b^{\prime}}_{l=1,\Lambda_{1}=([1,1],[1])}
&=&
\frac{1}{4}\delta_{ac}
(\delta_{bb^{\prime}}\delta_{a^{\prime}c^{\prime}}
-\delta_{ba^{\prime}}\delta_{b^{\prime}c^{\prime}})
+\frac{1}{4}\delta_{bc}
(
\delta_{aa^{\prime}}\delta_{b^{\prime}c^{\prime}}
-\delta_{ab^{\prime}}\delta_{a^{\prime}c^{\prime}}
), 
\end{eqnarray}
giving rise to the following 
covariant operators 
\begin{eqnarray}
O_{[2][1],l=0}&=&
\frac{1}{2}
(X_{a}\otimes  X_{b}\otimes X_{c}^{\ast}
+X_{b}\otimes  X_{a}\otimes X_{c}^{\ast})
-O_{l=1,\Lambda_{1}=([2][1])}
\cr
O_{[1,1][1],l=0}&=&
\frac{1}{2}
(X_{a}\otimes  X_{b}\otimes X_{c}^{\ast}
-X_{b}\otimes  X_{a}\otimes X_{c}^{\ast})
-O_{l=1,\Lambda_{1}=([1,1][1])}
\cr
O_{l=1,\Lambda_{1}=([2][1])}
&=&
\frac{1}{8}\delta_{ac}
(
X_{c^{\prime}}\otimes  X_{b}\otimes X_{c^{\prime}}^{\ast}
+X_{b}\otimes  X_{c^{\prime}}\otimes X_{c^{\prime}}^{\ast})
\cr
&&
+\frac{1}{8}\delta_{bc}
(
X_{a}\otimes  X_{c^{\prime}}\otimes X_{c^{\prime}}^{\ast}
+X_{c^{\prime}}\otimes  X_{a}\otimes X_{c^{\prime}}^{\ast}
)
\cr
O_{l=1,\Lambda_{1}=([1,1][1])}
&=&
\frac{1}{4}\delta_{ac}
(
X_{c^{\prime}}\otimes  X_{b}\otimes X_{c^{\prime}}^{\ast}
-X_{b}\otimes  X_{c^{\prime}}\otimes X_{c^{\prime}}^{\ast})
\cr
&&
+\frac{1}{4}\delta_{bc}
(
X_{a}\otimes  X_{c^{\prime}}\otimes X_{c^{\prime}}^{\ast}
-X_{c^{\prime}}\otimes  X_{a}\otimes X_{c^{\prime}}^{\ast}
). 
\end{eqnarray}

We next present gauge invariant operators. 
For the $k=0$ representation of $\gamma$, 
some labels in (\ref{gaugeinvariant}) are suppressed to give  
\begin{eqnarray}
{\cal O}_{\Lambda}^{\gamma (k=0)}
&=&
\sum_{b\in B_{N}(2,1)}
D^{\gamma}(b^{\ast})
tr_{2,1}(bO_{\Lambda})
=\frac{1}{t^{\gamma}}
tr_{2,1}(P^{\gamma}O_{\Lambda}), 
\end{eqnarray}
where $P^{\gamma}$ is the central projection operator 
in the Brauer algebra, which is given in \cite{0709.2158} by 
\begin{eqnarray}
&&P_{[2]\bar{[1]}}
=
\left(1-\frac{1}{N+1}C\right)p_{[2]},  \cr
&&P_{[1^{2}]\bar{[1]}}
=
\left(1-\frac{1}{N-1}C\right)p_{[1^{2}]}.
\label{centralprojector21}
\end{eqnarray}
Here
 $C:= C_{1\bar{1}}+C_{2\bar{1}}$ commutes with any element in 
$ \mathbb C (S_2)$. 
For the $k=1$ representation, we have 
\begin{eqnarray}
{\cal O}_{\Lambda,\Lambda_{1}}
^{\gamma (k=1),A}
&=&
\sum_{b\in B_{N}(2,1)}
D^{\gamma}_{A}(b^{\ast})
tr_{2,1}(bO_{\Lambda,\Lambda_{1}})
=\frac{1}{t^{\gamma}}
tr_{2,1}(P^{\gamma}_{A}O_{\Lambda,\Lambda_{1}}),
\end{eqnarray}
where
\begin{eqnarray}
&&
P_{[2]\bar{[1]}}^{ ( k=1 , \gamma_{+}=[1],\gamma_{-}=\emptyset)}
=\frac{1}{N+1}Cp_{[2]}, \cr
&&
P_{[1^{2}]\bar{[1]}}^{(k=1 , \gamma_{+}=[1],\gamma_{-}=\emptyset) }
=\frac{1}{N-1}Cp_{[1^{2}]} . 
\label{symmetricprojector21}
\end{eqnarray}
We exhibit some of them explicitly 
\begin{eqnarray}
&&tr_{2,1}(P_{[2]\bar{[1]}}O_{[2][1],l=0})
\cr
&=&
\frac{1}{2}\left(
trX_{a}trX_{b}trX_{c}^{\dagger}
+tr(X_{a}X_{b})trX_{c}^{\dagger}
\right)
\cr
&& 
-\frac{1}{2}\frac{1}{N+1}\left(
trX_{a}tr(X_{b}X_{c}^{\dagger})+trX_{b}tr(X_{a}X_{c}^{\dagger})
+tr(X_{a}X_{b}X_{c}^{\dagger})+tr(X_{b}X_{a}X_{c}^{\dagger})
\right)
\cr
&&
-tr_{2,1}(P_{[2]\bar{[1]}}O_{l=1,\Lambda_{1}=[2][1]}),
\cr
&&
\cr
&&
tr_{2,1}(P_{[2]\bar{[1]}}O_{l=1,\Lambda_{1}=[2][1]})
\cr
&=&
\frac{1}{8}\delta_{ac}
(
trX_{b}tr X_{c^{\prime}}trX_{c^{\prime}}^{\dagger}
+tr(X_{b}X_{c^{\prime}})trX_{c^{\prime}}^{\dagger}
)
\cr
&&
+\frac{1}{8}
\delta_{bc}
(trX_{a}tr X_{c^{\prime}}trX_{c^{\prime}}^{\dagger}
+tr(X_{a}X_{c^{\prime}})trX_{c^{\prime}}^{\dagger})
\cr
&&
-
\frac{1}{8}\frac{1}{N+1}
\delta_{ac}
(trX_{b}tr(X_{c^{\prime}} X_{c^{\prime}}^{\dagger})
+tr X_{c^{\prime}}tr( X_{b} X_{c^{\prime}}^{\dagger})
+tr (X_{b} X_{c^{\prime}} X_{c^{\prime}}^{\dagger})
+tr (X_{b} X_{c^{\prime}}^{\dagger} X_{c^{\prime}})
)
\cr
&&
-
\frac{1}{8}\frac{1}{N+1}
\delta_{bc}
(trX_{a}tr(X_{c^{\prime}} X_{c^{\prime}}^{\dagger})
+tr X_{c^{\prime}}tr( X_{a} X_{c^{\prime}}^{\dagger})
+tr (X_{a} X_{c^{\prime}} X_{c^{\prime}}^{\dagger})
+tr (X_{a}  X_{c^{\prime}}^{\dagger}X_{c^{\prime}})
). 
\nonumber
\end{eqnarray}


\section{Quantum mechanics and 
conserved charges measuring representation labels
}
\label{sec:qmcasimirs}
In this section, we study the matrix 
quantum mechanics which is obtained by the dimensional reduction 
of the four-dimensional ${\cal N}=4$ SYM on $S^{3}\times R$. 
Because we are mainly interested in the free theory, 
we ignore the interaction terms. 
It is found that the Hamiltonian can be given by a set of 
the harmonic oscillators \cite{0111222}, and 
it has been known that the harmonic oscillator 
is characterised by many conserved charges. 
Some conserved charges 
whose eigenvalues can specify an orthogonal state 
were constructed in \cite{0807.3696}. 
Furthermore, 
the construction was based on symmetries 
which are enhanced at the free level. 
In this section, we show conserved charges 
for the present context. 

The Hamiltonian and $U(1)^{3}$ charge are 
\begin{eqnarray}
&&H=\sum_{a}tr(A_{a}^{\dagger}A_{a}+B_{a}^{\dagger}B_{a})+N^{2}, \cr
&&J_{a}=tr(B_{a}^{\dagger}B_{a}-A_{a}^{\dagger}A_{a}),
\end{eqnarray}
where we have introduced the matrix annihilation and creation operators
with the non-zero commutation relations
\begin{eqnarray}
&&[(A_{a})_{ij},(A^{\dagger}_{b})_{kl}]=
\delta_{ab}\delta_{jk}\delta_{il}, \quad
[(B_{a})_{ij},(B^{\dagger}_{b})_{kl}]=
\delta_{ab}\delta_{jk}\delta_{il}. 
\end{eqnarray}
Gauge covariant states are obtained by acting with $A^{\dagger}$ 
and $B^{\dagger}$ on the vacuum with $A|0\rangle =B|0\rangle =0$. 
Because the gauge invariance enforces all upper indices 
to be contracted with all lower indices, 
all gauge invariant operators take the form of 
$tr_{m,n}(\Sigma(b)(A^{\dagger})^{ \otimes m}\otimes
( B^{\dagger})^{ \otimes n})$. 
Here 
$\Sigma$ is a map from elements in $B_{N}(m,n)$ to elements in
$\mathbb C(S_{m+n})$ introduced in \cite{0709.2158}\footnote{
One simple example is 
$tr(XX^{\dagger})=tr_{1,1}(CX\otimes X^{\ast})=
tr_{2}(\Sigma(C)X\otimes X^{\dagger})$.  
The map $\Sigma$ was also exploited in \cite{0801.2061} to 
construct another non-holomorphic extension for a single complex 
matrix.  
}.
Our claim in the previous sections can be straightforwardly applied to 
give the following orthogonal state, 
\begin{eqnarray}
&&|{\cal O}_{\Lambda,M_{\Lambda},\Lambda_{1},\tau}
^{\gamma,A,ij,\tau_{\Lambda_{1}}}\rangle
\cr
&&
\hspace{-0.4cm}=
\sum_{m_{A},m_{A}^{\prime},m_{\Lambda_{1}}}
C^{\tau_{\Lambda_{1}},\Lambda_{1},m_{\Lambda_{1}}}
_{A,m_{A},A,m_{A}^{\prime}}
\sum_{b}
D^{\gamma}_{A,m_{A},j,m_{A}^{\prime},i}(b^{\ast})
\sum_{\vec{a},\vec{b}}C_{\Lambda,M_{\Lambda},
\Lambda_{1},m_{\Lambda_{1}},\tau}^{\vec{a},\vec{b}}
tr_{m,n}(\Sigma(b)O_{\vec{a},\vec{b}})
|0\rangle, 
\end{eqnarray}
where 
\begin{eqnarray}
O_{\vec{a},\vec{b}}=
A_{a_{1}}^{\dagger}
\otimes \cdots \otimes 
A_{a_{m}}^{\dagger}
\otimes 
B_{b_{1}}^{\dagger}
\otimes \cdots \otimes 
B_{b_{n}}^{\dagger}. 
\end{eqnarray}
The inner product is diagonal:
\begin{eqnarray}
\langle 
{\cal O}_{\Lambda^{\prime},M_{\Lambda^{\prime}}^{\prime},\Lambda_{1}^{\prime},
\tau^{\prime}}
^{\gamma^{\prime},A^{\prime},i^{\prime}j^{\prime},\tau_{\Lambda_{1}}^{\prime}}
|
{\cal O}_{\Lambda,M_{\Lambda},\Lambda_{1},\tau}
^{\gamma,A,ij,\tau_{\Lambda_{1}}}\rangle
=m!n!d_{\Lambda_{1}}
\frac{1}{t^{\gamma}}
\delta_{\gamma\gamma^{\prime}}
\delta_{AA^{\prime}}
\delta_{\Lambda \Lambda^{\prime}}
\delta_{M_{\Lambda}M_{\Lambda^{\prime}}^{\prime}}
\delta_{\Lambda_{1}\Lambda_{1}^{\prime}}
\delta_{
\tau_{\Lambda_{1}}
\tau_{\Lambda_{1}^{\prime}}^{\prime}}
\delta_{\tau\tau^{\prime}} 
\delta_{ii^{\prime}}\delta_{jj^{\prime}}.
\end{eqnarray}

\vspace{0.4cm}

Before showing 
a set of conserved charges 
whose eigenvalues identify the labels of the orthogonal state, 
we shall explain 
symmetries of the Hamiltonian. 

The original scalar matrix field theory action is invariant 
under the adjoint unitary transformation. 
We can express the conserved matrix current 
in terms of the annihilation and creation operators as 
$[A_{a},A_{a}^{\dagger}]$ 
for the $A$-sector and 
$[B_{a},B_{a}^{\dagger}]$ 
for the $B$-sector.
We will find that 
we have more symmetries at the free level. 
In order to see enhanced symmetries of this Hamiltonian, 
we decompose\footnote{
Such decomposition was also considered in \cite{0607033}. 
} the generator and name them
as 
\begin{eqnarray}
&& (G_{L,A})_{ij}:= \sum_{a}(G_{L,A_{a}})_{ij}, \quad 
(G_{L,A_{a}})_{ij}:=(A_{a}^{\dagger})_{kj}(A_{a})_{ik}
\cr
&&
(G_{R,A})_{ij}:= \sum_{a}(G_{R,A_{a}})_{ij},\quad
(G_{R,A_{a}})_{ij}:=(A_{a}^{\dagger})_{ik}(A_{a})_{kj}
\end{eqnarray}
and similarly for the $B$-sector. 
We also define
\begin{eqnarray}
&& (G_{L})_{ij}:= (G_{L,A})_{ij}+(G_{L,B})_{ij}, 
\cr
&&
(G_{R})_{ij}:= ( G_{R,A})_{ij}+(G_{R,B})_{ij}. 
\end{eqnarray}
They generate 
the left action and the right action:
\begin{eqnarray}
&&[tr(\Lambda G_{L,A}),(A_{a})_{ij}]=-(\Lambda A_{a})_{ij}, 
\cr
&&
[tr(\Lambda G_{R,A}),(A_{a})_{ij}]=(A_{a}\Lambda )_{ij}. 
\end{eqnarray}
It is easy to show that 
\begin{eqnarray}
[G_{L,A},tr(A_{a}A_{a}^{\dagger})]=0, \quad
[G_{R,A},tr(A_{a}A_{a}^{\dagger})]=0,  
\end{eqnarray}
and similar equations for the $B$-sector, 
implying that all of 
$G_{L,A}$, $G_{R,A}$ and $G_{L,B}$, $G_{R,B}$ 
generate 
symmetries of the Hamiltonian. 
They form the $u(N)$ commutation relations 
\begin{eqnarray}
&& [(G_{L,A_{a}})_{ij},(G_{R,A_{a}})_{kl}]=0, \cr
&& 
[(G_{L,A_{a}})_{ij},(G_{L,A_{b}})_{kl}]=
\delta_{ab}
((G_{L,A_{a}})_{kj}\delta_{il}-(G_{L,A_{a}})_{il}\delta_{jk}), 
\cr
&&
[(G_{R,A_{a}})_{ij},(G_{R,A_{b}})_{kl}]=
\delta_{ab}
(
(G_{R,A_{a}})_{il}\delta_{jk}
-(G_{R,A_{a}})_{kj}\delta_{il}). 
\end{eqnarray}
Since we also have the same relations for the $B$-sector, 
we thus have four commuting copies of the $u(N)$ algebra. 

In terms of these generators, the Hamiltonian 
and the angular momentum can be expressed by
\begin{eqnarray}
&&
H=tr(G_{L,A}+G_{L,B}), 
\cr
&&
J_{a}=tr(G_{L,B_{a}}-G_{L,A_{a}}). 
\end{eqnarray}
It is mentioned that $tr(G_{L,A_{a}})=tr(G_{R,A_{a}})$ and 
$tr(G_{L,B_{a}})=tr(G_{R,B_{a}})$. 
Because 
$G_{L}+G_{R}$ generates the adjoint gauge transformation, 
we have 
\begin{eqnarray}
G_{L}+G_{R}=0
\end{eqnarray}
on gauge invariant states. 

\vspace{0.2cm}

From now on we present some conserved operators 
which act on the orthogonal state with 
eigenvalues 
measuring the representation labels. 
We shall build operators from
the symmetry generators, so  
it is manifest that they commute with the Hamiltonian. 
The construction of those operators is shown in \cite{0807.3696}. 
We define 
the symbol $\doteq$ to assume 
the actions on 
$|{\cal O}_{\Lambda,M_{\Lambda},\Lambda_{1},\tau}
^{\gamma,A,ij,\tau_{\Lambda_{1}}}\rangle$. 
We first show three operators 
\begin{eqnarray}
&&
tr(G_{L})^{2}
\doteq tr(G_{R})^{2} \doteq
C_{2}(\gamma),
\cr
&&
tr\left(G_{L,A}\right)^{2}
\doteq 
tr\left(G_{R,A}\right)^{2} \doteq 
 C_{2}(\alpha),
\cr
&&
tr\left(G_{L,B}\right)^{2}
\doteq 
tr\left(G_{R,B}\right)^{2}
\doteq 
C_{2}(\beta), 
\end{eqnarray}
where $C_{2}$ is the quadratic Casimir of $U(N)$. 
These commute each other without assuming the action on the state. 
We also have similar equations for higher order actions, 
for example, 
\begin{eqnarray}
tr\left(G_{L}\right)^{r}\doteq 
tr\left(G_{R}\right)^{r}\doteq 
C_{r}(\gamma). 
\end{eqnarray}
It was also shown that the multiplicity associated 
with the decomposition $\gamma \rightarrow A$ 
of the colour structure 
can be measured
as 
\begin{eqnarray}
&&
tr\left(\left(G_{L,A}\right)^{2}
G_{L,B}\right)
\doteq 
C(\gamma,\alpha,\beta,i),
\cr
&&
tr\left(\left(G_{R,A}\right)^{2}
G_{R,B}\right)
\doteq C(\gamma,\alpha,\beta,j),
\end{eqnarray}
where $C(\gamma,\alpha,\beta,i)$ is a quantity 
depending on $\gamma$, $\alpha$, $\beta$, $i$, but 
the exact form has not been found
\footnote{
In the paper \cite{0807.3696}, 
it was shown that 
$tr\left(\left(G_{L,A}\right)^{2}G_{L,B}\right)$ 
can recognise the multiplicity index, and 
it was confirmed for some examples.}.

In addition to the enhanced symmetry generators 
$G_{L,A}$, $G_{R,A}$ and $G_{L,B}$, $G_{R,B}$, we have 
another enhanced symmetry which is generated by 
\begin{eqnarray}
(G_{E})_{kljm}:=(G_{E,A})_{kljm}+(G_{E,B})_{kljm}, 
\end{eqnarray}
where 
\begin{eqnarray}
(G_{E,A})_{kljm}:=\sum_{a}(G_{E,A_{a}})_{kljm}
:=\sum_{a}(A_{a})_{kl}^{\dagger}(A_{a})_{jm}. 
\end{eqnarray}
We find that this generates a $U(N^{2})$ symmetry, 
which can be manifested by introducing 
a composite index $I=(i,j)$, where $I$ takes $N^{2}$ values when 
$i$ and $j$ run over $N$ values. 
The use of the composite index enables us to express 
the generator as 
\begin{eqnarray}
(G_{E,A})_{IJ}
=\sum_{a}(A_{a})_{I}^{\dagger}(A_{a})_{J}. 
\end{eqnarray}
The commutation relation of $G_{E}$ is 
\begin{eqnarray}
[(G_{E})_{IJ},(G_{E})_{KL}]
=(G_{E})_{IL}\delta_{JK}-(G_{E})_{KJ}\delta_{IL}.
\end{eqnarray}
The actions of $G_{E,A}$ on $A_{a}$ and $A^{\dagger}_{a}$ 
are 
\begin{eqnarray}
&&
[(G_{E,A})_{IJ},(A_{b}^{\dagger})_{K}]
=\delta_{JK}(A_{b}^{\dagger})_{I},
\cr
&&
[(G_{E,A})_{IJ},(A_{b})_{K}]
=-\delta_{IK}(A_{b})_{J}, 
\end{eqnarray}
where we have defined 
$\delta_{JK}=\delta_{j_{1}k_{2}}\delta_{j_{2}k_{1}}$ for 
$J=(j_{1},j_{2})$ and $K=(k_{1},k_{2})$.  
We also have 
\begin{eqnarray}
&&
[(G_{E,A})_{kljm},(G_{L,A})_{pq}]
=(G_{E,A})_{klpm}\delta_{jq}-(G_{E,A})_{kqjm}\delta_{lp}, 
\cr
&&
[(G_{E,A})_{kljm},(G_{R,A})_{pq}]
=(G_{E,A})_{kljq}\delta_{pm}-(G_{E,A})_{pljm}\delta_{qk}. 
\end{eqnarray}
It is also easy to see that 
the symmetry generated by 
$G_{E}$ is indeed a symmetry of the Hamiltonian:
\begin{eqnarray}
[(G_{E,A})_{kljm},
tr(A_{a}^{\dagger}A_{a})]
=0.
\end{eqnarray}
With the enhanced $U(N^{2})$ symmetry 
we can 
define an operator which can measure 
$\tau_{1}$, i.e. 
the number of copies of $\Lambda_{1}$ in the inner tensor product 
of $A\otimes A$, as 
\begin{eqnarray}
tr\left(G_{L}G_{E}G_{R}\right)\doteq C(\tau_{\Lambda_{1}}). 
\end{eqnarray}
$C(\tau_{\Lambda_{1}})$ is a quantity depending on $\tau_{1}$. 

The representation labels associated with the flavour indices 
can be measured in terms of $E_{A}{}_{ab}:=tr(A_{a}^{\dagger}A_{b})$, 
whose action on $A^{\dagger}$ is 
\begin{eqnarray}
[E_{A}{}_{ab}, (A_{c}^{\dagger})_{ij}]
=\delta_{bc}(A_{a}^{\dagger})_{ij}. 
\end{eqnarray}
$E_{A}{}_{ab}$ satisfies the $u(p)$ commutation relation: 
\begin{eqnarray}
[E_{A}{}_{ab},E_{A}{}_{cd}]
=\delta_{bc}E_{A}{}_{ad}
-\delta_{ad}E_{A}{}_{cb}. 
\end{eqnarray}
The quadratic Casimir of $u(p)$ appears an eigenvalue of 
the quadratic action of $E_{ab}:=E_{A}{}_{ab}+E_{B}{}_{ab}$ 
on the orthogonal state: 
\begin{eqnarray}
E_{ab}E_{ab} \doteq C_{2}(\Lambda). 
\end{eqnarray}
We also have 
\begin{eqnarray}
E_{Aab}E_{Aab} \doteq C_{2}(\alpha_{1}), \quad  
E_{Bab}E_{Bab} \doteq C_{2}(\beta_{1})
\end{eqnarray}
for $\Lambda_{1}=(\alpha_{1},\beta_{1})$. 
The index $\tau$ which runs over the multiplicity of $\Lambda_{1}$ 
in $\Lambda$ would be measured by 
\begin{eqnarray}
E_{Aab}E_{Abc}E_{Bca}.   
\end{eqnarray}

\vspace{0.4cm}

In summary, we have provided 
some conserved operators 
which can be diagonalised by the orthogonal state. 
In other words, the orthogonal state proposed in this paper 
can be specified by the simultaneous diagonalisation of 
the conserved operators. 

Operators which have the diagonal actions on 
an orthogonal state can be 
organised by 
the enhanced symmetries of the Hamiltonian. 
General multi-trace operators made from those 
generators are all conserved, and they would form 
an extension of the $W_{\infty}$ algebra. 
It is interesting to ask how the enhanced symmetries 
can be connected with 
the integrability of this system. 
Conserved charges may also be helpful to 
understand gravitational duals or space-time interpretations 
(see \cite{0602263, 0705.4431} for such discussions).


\section{Discussions}
\label{sec:discussion}
In this paper, we have 
proposed an orthogonal set of 
non-holomorphic gauge invariant 
operators made from some complex matrices 
at the free coupling 
based on the Brauer algebra. 
Below are possible future directions along this line.
 
\vspace{0.2cm}

One remaining problem which should be discussed is 
to see if this proposed 
operator exhausts multi-matrix gauge invariant operators.
For the case of a single complex matrix, 
a counting formula in terms of group theoretic languages 
was given in \cite{0709.2158}, 
and it is proved for large $N$ in \cite{YKSRDT}. 
It has not, however, been cleared 
how the counting is modified when $N$ is finite 
in the context of Brauer algebra. 
One modification of the finite $N$ case is to impose the constraint 
$c_{1}(\gamma_{+})+c_{1}(\gamma_{-}) \le N$, which is naturally 
expected from the definition of the $U(N)$ group. 
This constraint seems to hold the attention because 
it realises a generalisation of the cut-off $c_{1}(R)\le N$ 
for the half-BPS case. 
In the half-BPS case, the cut-off can be translated into 
the cut-off for the angular momentum of giant gravitons \cite{0003075}. 
(Such an effect was originally studied in \cite{9804085,9902059}.)
The cut-off for the present case would give 
a constraint for 
a composite angular momentum of giant gravitons 
and anti-giant gravitons.  
For the single complex matrix case, 
more analyses at finite $N$ will be given in 
\cite{YKSRDT}. 
The counting operators including the finite $N$ case 
for the non-holomorphic multi-matrix case
will be discussed in future publications. 

\vspace{0.2cm}

Another basis of non-holomorphic one-matrix 
gauge invariant operators 
was built in \cite{0801.2061}, where
the symmetric group plays a role instead of the Brauer algebra. 
This basis was originally proposed to study excitations of 
giant gravitons \cite{0411205,0701066,0701067,0710.5372}. 
It will be possible to construct another basis 
for non-holomorphic multi-matrix operators as 
an extension of \cite{0801.2061}. 
As is discussed in \cite{0807.3696}, 
a difference of two bases in a sector 
can be explained by the fact that 
a set of Casimirs 
characterising 
an orthogonal basis 
does not commute with 
another set of Casimirs characterising another 
orthogonal basis. 
It would be nice to 
ask roles of two different bases from the point of view of 
dual physics. 

\vspace{0.2cm}

For the highest weight state in 
the half-BPS sector, the physics is characterised by 
a single Young diagram alone. 
The number and the angular momentum of 
giant gravitons are encrypted in a single Young diagram. 
On the other hand, 
the Brauer algebra brings in 
two kinds of representation labels, i.e. 
$\gamma=(\gamma_{+},\gamma_{-}, k)$, and 
$A=(\alpha, \beta)$. 
Having diagonalised two-point functions at the free level, 
the representation labels will be directly related to 
a tensionless string theory.  
In our previous paper \cite{0709.2158}, 
it was conjectured that the operator has a close connection 
with a system of giant gravitons and anti-giant gravitons. 
The $k=0$ representation 
would be naturally related to such a system, 
where 
$\gamma_{+}=\alpha$ would read 
the number and angular momentum of 
giant gravitons while $\gamma_{-}=\beta$ would read those of 
anti-giant gravitons. 
Because $k$ is the number of boxes 
which are got rid of from $m$ boxes and $n$ anti-boxes, 
one can anticipate that 
$k\neq 0$ representations 
would contain branes and anti-branes 
with smaller quantum numbers. 
At $k=m=n$, one may expect closed string excitations without 
branes and anti-branes. 
If we include corrections of the t'Hooft coupling, 
this system may start showing the instability originated from 
the existence of branes and anti-branes. 
If our operators can really describe physics of 
the unstable system, it is natural to expect physics of 
tachyon condensation to be encoded into the representation labels 
in a way. 
Fortunately, 
it is possible to study this system at finite string scale 
where tachyon has the negative mass squared
because 
this $SO(6)$ sector is closed at one-loop. 
In this sense, this system will be a good framework 
to know how the Brauer algebra captures such an interesting 
unstable system.  
The operator mixing problem 
has been 
reported in \cite{0205321,0206020,0208178,0209002,0312228}. 
On the other hand, 
the one-loop correction was studied in the language of 
the representation basis for the $U(2)$ holomorphic sector 
in \cite{0801.2094}, where 
a restricted mixing pattern was found for Young diagrams. 
A similar restricted mixing was also given in 
\cite{0701067,0710.5372} in the context of the restricted Schur polynomials. 
The mixing problem was studied in \cite{0404066} in terms 
of a basis expressed by the symmetric group, and 
the Hamiltonian was expressed by splitting and joining interactions. 
Studies along these lines using the bases proposed in this paper 
would shed light on the role of 
the Brauer algebra.  

\vspace{0.2cm}

Our orthogonal operators are entirely specified by 
some group theoretic operators 
constructed from generators for the enhanced symmetries.  
Because Casimir operators know 
information about 
representation labels of orthogonal sets, studies of 
Casimirs operators at the one-loop level would tell us 
about the operator mixing. For example, a relation 
between the one-loop dilatation operator and the Casimirs 
may suggest how orthogonal operators mix under 
quantum corrections. 
The breaking of the enhanced symmetries 
with the interactions turned on 
can be possibly associated with the operator mixing problem. 
We hope to understand the role of the enhanced symmetries 
by looking for a connection to the integrability.

\vspace{1cm}

\begin{center}
{\bf \large Acknowledgements}
\end{center}

I would like to thank Sanjaye Ramgoolam for valuable discussions 
and for collaboration at the initial stage of this work, 
and 
Tom Brown, Robert de Mello Koch, 
Paul Heslop, David Turton for useful discussions. 
I gratefully acknowledge the support of STFC grant PP/D507323/1. 
I also thank Okayama Institute for Quantum Physics for hospitality, 
where the final part of this work was done. 
Special thanks to Sea the Stars 
and Liverpool FC for giving me encouragement.   

\vspace{1cm}


\renewcommand{\theequation}{\Alph{section}.\arabic{equation}}
\appendix

\section{One-loop dilatation operator}
\setcounter{equation}{0} 
\label{sec:hamiltoniancomplex}
In this section, we rewrite the one-loop dilatation operator in 
terms of the complex variables.  

The dilatation operator in the $SO(6)$ sector 
up to one-loop order 
was given \cite{Beisert:2003jj,Beisert:2004ry} by 
\begin{eqnarray}
{\cal D}={\cal D}_{0}+{\cal D}_{2}
:={\cal D}_{0}+g^{2}{\cal H}, \qquad (g^{2}=g_{YM}^{2}N/8\pi^{2})
\end{eqnarray}
where 
\begin{eqnarray}
{\cal H}=N^{-1}\left(
-\frac{1}{2}:tr[\Phi_{m},\Phi_{n}][\check{\Phi}^{m},\check{\Phi}^{n}]:
-\frac{1}{4}:tr[\Phi_{m},\check{\Phi}^{n}]
[\Phi_{m},\check{\Phi}^{n}]:
\right). 
\label{hamiltonian}
\end{eqnarray}
$\Phi_{m}$ $(m=1,\dots,6)$ is the scalar fields.  
$\check{\Phi}$ represents the following derivative action
\begin{eqnarray}
(\check{\Phi}^{m})_{ij}(\Phi_{n})_{kl}=\delta_{n}^{m}
\delta_{il}\delta_{kj}
\label{derivativeonUN}
\end{eqnarray}
for the $u(N)$ gauge group. 

We define the complex combination of two scalars
\begin{eqnarray}
&& X_{a} = \Phi_{2a-1}+i\Phi_{2a}, \quad (a=1,2,3). 
\end{eqnarray}
Solving for $\Phi$, we get
\begin{eqnarray}
\Phi_{2a-1} =\frac{1}{2}\left( X_{a}+X_{a}^{\dagger}\right),\quad
\Phi_{2a} = 
\frac{1}{2i}\left(
X_{a}-X_{a}^{\dagger}\right).
\end{eqnarray}
We express these equations for later convenience as 
\begin{eqnarray}
\Phi_{m}=a_{m}^{i}Z_{i},
\end{eqnarray}
where 
$Z_{i}=(X_{1},X_{1}^{\dagger},X_{2},X_{2}^{\dagger},X_{3},X_{3}^{\dagger})$. 
The derivative of the complex matrix is defined by 
\begin{eqnarray}
 \check{X}^{a} = 
\frac{1}{2}\left(\check{\Phi}^{2a-1}-i\check{\Phi}^{2a}\right), 
\end{eqnarray}
satisfying  
$(\check{X}^{a})_{ij}(X_{b})_{kl}=\delta_{ab}\delta_{il}\delta_{kl}$. 
We solve for $\check{\Phi}$ as 
\begin{eqnarray}
\check{\Phi}_{2a-1} = 
\check{X}_{a}+\check{X}_{a}^{\dagger},\quad
\check{\Phi}_{2a} = 
i\left(
\check{X}_{a}-\check{X}_{a}^{\dagger}\right),
\end{eqnarray}
which 
we shall denote compactly by 
\begin{eqnarray}
\check{\Phi}_{m}=b_{m}^{i}\check{Z}_{i}, 
\end{eqnarray}
where 
$\check{Z}_{i}=(\check{X}_{1},\check{X}_{1}^{\dagger},
\check{X}_{2},\check{X}_{2}^{\dagger},
\check{X}_{3},\check{X}_{3}^{\dagger})$. 

Using these complex variables, we can rewrite the first term of 
the Hamiltonian (\ref{hamiltonian}) as  
\begin{eqnarray}
tr[\Phi_{m},\Phi_{n}][\check{\Phi}^{m},\check{\Phi}^{n}] 
&=&a_{m}^{i}a_{n}^{j}b_{m}^{k}b_{n}^{l}
tr[Z_{i},Z_{j}][\check{Z}^{k},\check{Z}^{l}] \cr
&=&g^{ik}g^{jl}
tr[Z_{i},Z_{j}][\check{Z}^{k},\check{Z}^{l}] \cr
&=&
tr[X_{a},X_{b}][\check{X}_{a},\check{X}_{b}]
+tr[X_{a},X_{b}^{\dagger}][\check{X}_{a},\check{X}^{\dagger}_{b}]\cr
&& 
+tr[X_{a}^{\dagger},X_{b}][\check{X}_{a}^{\dagger},\check{X}_{b}]
+tr[X_{a}^{\dagger},X_{b}^{\dagger}]
[\check{X}_{a}^{\dagger},\check{X}_{b}^{\dagger}].
\label{rewritefirst}
\end{eqnarray}
We have defined the metric $g^{ik}=\sum_{m}a_{m}^{i}b_{m}^{k}=\delta^{ik}$.
On the other hand, 
the second term of the Hamiltonian 
can be rewritten as 
\begin{eqnarray}
tr[\Phi_{m},\check{\Phi}^{n}]
[\Phi_{m},\check{\Phi}^{n}] 
&=&
a_{m}^{i}b_{n}^{j}a_{m}^{k}b_{n}^{l}
tr[Z_{i},\check{Z}^{j}]
[Z_{k},\check{Z}^{l}]
\cr
&=&h^{ik}\tilde{h}^{jl}
tr[Z_{i},\check{Z}^{j}][Z_{k},\check{Z}^{l}] \cr
&=&
tr[X_{a},\check{X}_{b}][X_{a}^{\dagger},\check{X}_{b}^{\dagger}]
+tr[X_{a},\check{X}^{\dagger}_{b}][X_{a}^{\dagger},\check{X}_{b}]\cr
&&
+tr[X_{a}^{\dagger},\check{X}^{\dagger}_{b}][X_{a},\check{X}_{b}]
+tr[X_{a}^{\dagger},\check{X}_{b}][X_{a},\check{X}_{b}^{\dagger}] \cr
&=&
2tr[X_{a},\check{X}_{b}][X_{a}^{\dagger},\check{X}_{b}^{\dagger}]
+2tr[X_{a},\check{X}^{\dagger}_{b}][X_{a}^{\dagger},\check{X}_{b}],
\label{rewritesecond}
\end{eqnarray}
where we have defined the metric as 
$h^{ik}=a_{m}^{i}a_{m}^{k}$ and 
$\tilde{h}^{jl}=b_{n}^{j}b_{n}^{l}$, 
with non-zero components  
$h^{X_{a}\bar{X}_{a}}=1/2$ and $\tilde{h}^{X_{a}\bar{X}_{a}}=2$. 
Collecting (\ref{rewritefirst}) and (\ref{rewritesecond}), 
we get the Hamiltonian in terms of the complex variables as 
\begin{eqnarray}
{\cal H}&=&
-\frac{1}{2N}:\left(
tr[X_{a},X_{b}][\check{X}_{a},\check{X}_{b}]
+tr[X_{a},X_{b}^{\dagger}][\check{X}_{a},\check{X}^{\dagger}_{b}]
\right.
\cr
&& \left.
+tr[X_{a}^{\dagger},X_{b}][\check{X}_{a}^{\dagger},\check{X}_{b}]
+tr[X_{a}^{\dagger},X_{b}^{\dagger}]
[\check{X}_{a}^{\dagger},\check{X}_{b}^{\dagger}]  
\right.
\cr
&& \left.
+tr[X_{a},\check{X}_{b}][X_{a}^{\dagger},\check{X}_{a}^{\dagger}]
+tr[X_{a},\check{X}^{\dagger}_{b}][X_{a}^{\dagger},\check{X}_{b}]
\right):.
\end{eqnarray}


\section{Proof of the diagonal two-point function}
\label{proofdiagonaltwopoint}
\setcounter{equation}{0} 
In this section, we show detailed calculations 
of the two-point functions. 

\subsection{Two-point function for the covariant operator}
\label{proofdiagonaltwopointcovariant}
The two-point function of the representation basis 
(\ref{operatorrepbasis}) can be computed as 
\begin{eqnarray}
&&\langle 
: (O_{\Lambda,M_{\Lambda},\Lambda_{1},m_{\Lambda_{1}},
\tau}^{\dagger})_{J}^{I}:
: (O_{\Lambda^{\prime},M_{\Lambda^{\prime}}^{\prime},\Lambda_{1}^{\prime},
m_{\Lambda_{1}^{\prime}}^{\prime},\tau^{\prime}})_{L}^{K}:
\rangle \cr
&=&
\sum_{a,b,a^{\prime},b^{\prime}}
(C_{\Lambda,M_{\Lambda},
\Lambda_{1},m_{\Lambda_{1}},\tau}^{\vec{a},\vec{b}})^{\ast}
C_{\Lambda^{\prime},M_{\Lambda^{\prime}}^{\prime},
\Lambda_{1}^{\prime},m_{\Lambda_{1}^{\prime}}^{\prime},\tau^{\prime}}
^{\vec{a^{\prime}},\vec{b^{\prime}}}
\langle : (O_{\vec{a},\vec{b}}^{\dagger})_{J}^{I}:
: (O_{\vec{a^{\prime}},\vec{b^{\prime}}})_{L}^{K}:
\rangle \cr
&=&\sum_{a,b,a^{\prime},b^{\prime}}
(C_{\Lambda,M_{\Lambda},
\Lambda_{1},m_{\Lambda_{1}},\tau}^{\vec{a},\vec{b}})^{\ast}
C_{\Lambda^{\prime},M_{\Lambda^{\prime}}^{\prime},
\Lambda_{1}^{\prime},m_{\Lambda_{1}^{\prime}}^{\prime},\tau^{\prime}}
^{\vec{a^{\prime}},\vec{b^{\prime}}}
\sum_{\sigma \in S_{m}\times S_{n}}
\prod_{k=1}^{m}\delta_{a_{k}a^{\prime}_{\sigma(k)}} 
\prod_{l=1}^{n}\delta_{b_{l}b^{\prime}_{\sigma(l)}} 
(\sigma)_{J}^{K}(\sigma^{-1})_{L}^{I}\cr
&=&\sum_{a,b}
\sum_{\sigma \in S_{m}\times S_{n}}
(C_{\Lambda,M_{\Lambda},
\Lambda_{1},m_{\Lambda_{1}},\tau}^{\vec{a},\vec{b}})^{\ast}
C_{\Lambda^{\prime},M_{\Lambda^{\prime}}^{\prime},
\Lambda_{1}^{\prime},m_{\Lambda_{1}^{\prime}}^{\prime},\tau^{\prime}}
^{a_{\sigma^{-1}(1)},\cdots,a_{\sigma^{-1}(m)},
b_{\sigma^{-1}(1)},\cdots,b_{\sigma^{-1}(n)}}
(\sigma)_{J}^{K}(\sigma^{-1})_{L}^{I}\cr
&=&
\delta_{\Lambda \Lambda^{\prime}}
\delta_{M_{\Lambda}M_{\Lambda^{\prime}}^{\prime}}
\delta_{\Lambda_{1}\Lambda_{1}^{\prime}}
\delta_{\tau\tau^{\prime}}
\sum_{\sigma \in S_{m}\times S_{n}}
D^{\Lambda_{1}}_{m_{\Lambda_{1}}m_{\Lambda_{1}^{\prime}}^{\prime}}(\sigma)
(\sigma)_{J}^{K}(\sigma^{-1})_{L}^{I}.
\end{eqnarray}
To get the third equality, we solved the delta symbols. 
In the last step, we have used 
\begin{eqnarray}
C_{\Lambda^{\prime},M_{\Lambda^{\prime}}^{\prime},
\Lambda_{1}^{\prime},m_{\Lambda_{1}^{\prime}}^{\prime},\tau^{\prime}}
^{a_{\sigma^{-1}(1)},\cdots,a_{\sigma^{-1}(m)},
b_{\sigma^{-1}(1)},\cdots,b_{\sigma^{-1}(n)}}
=
D^{\Lambda_{1}}
_{m_{\Lambda_{1}^{\prime}}^{\prime}
m_{\Lambda_{1}^{\prime}}^{\prime\prime}}(\sigma^{-1})
C_{\Lambda^{\prime},M_{\Lambda^{\prime}}^{\prime},
\Lambda_{1}^{\prime},m_{\Lambda_{1}^{\prime}}^{\prime\prime},\tau^{\prime}}
^{a_{1},\cdots,a_{m},
b_{1},\cdots,b_{n}}
\label{clebshsigmainverse}
\end{eqnarray}
and 
the following equation 
\begin{eqnarray}
\sum_{\vec{a},\vec{b}}
(C_{\vec{a},\vec{b}}
^{\Lambda,M_{\Lambda},\Lambda_{1},m_{\Lambda_{1}},\tau})^{\ast}
C_{\vec{a},\vec{b}}^{\Lambda^{\prime},M_{\Lambda^{\prime}}
^{\prime},\Lambda_{1}^{\prime},m_{\Lambda_{1}^{\prime}}
^{\prime},\tau^{\prime}}
=\delta_{\Lambda \Lambda^{\prime}}
\delta_{M_{\Lambda}M_{\Lambda^{\prime}}^{\prime}}
\delta_{\Lambda_{1}\Lambda_{1}^{\prime}}
\delta_{m_{\Lambda_{1}}m_{\Lambda_{1}^{\prime}}^{\prime}}
\delta_{\tau\tau^{\prime}}. 
\label{propertyCG}
\end{eqnarray}
The derivation of this equation is 
completely similar to (69) in \cite{0806.1911}. 
(\ref{clebshsigmainverse}) comes from the fact that 
$\sigma^{-1}$ acts only on 
$V_{\Lambda_{1}}^{\mathbb C(S_{m}\times S_{n})}$ 
when 
it acts on $V^{\otimes m}\otimes \bar{V}^{\otimes n}$ 
in (\ref{scdecomposed}).


\subsection{Two-point function for the gauge invariant operator}

In this subsection, we shall present the proof of 
(\ref{diagonaltwopoint}). 
Using the two-point function of the covariant 
operator (\ref{covariant2pt}), we get
\begin{eqnarray}
&&\langle 
{\cal O}^{\gamma,A,ij,\tau_{\Lambda_{1}}}
_{\Lambda,M_{\Lambda},\Lambda_{1},\tau}{}^{\dagger}
{\cal O}^{\gamma^{\prime},A^{\prime},i^{\prime}j^{\prime},
\tau_{\Lambda_{1}^{\prime}}^{\prime}}
_{\Lambda^{\prime},M_{\Lambda^{\prime}}^{\prime},
\Lambda_{1}^{\prime},\tau^{\prime}}
\rangle \cr
&=&
\delta_{\Lambda \Lambda^{\prime}}
\delta_{M_{\Lambda}M_{\Lambda^{\prime}}^{\prime}}
\delta_{\Lambda_{1}\Lambda_{1}^{\prime}}
\delta_{\tau\tau^{\prime}} 
C^{\tau_{\Lambda_{1}},\Lambda_{1},m_{\Lambda_{1}}}
_{A,m_{A},A,m_{A}^{\prime}}
C^{\tau_{\Lambda_{1}^{\prime}}^{\prime},\Lambda_{1}
^{\prime},m_{\Lambda_{1}^{\prime}}^{\prime}}
_{A^{\prime},m_{A^{\prime}},A^{\prime},m_{A^{\prime}}^{\prime}}
\cr
&&
\hspace{-0.4cm}
\sum_{\sigma \in S_{m}\times S_{n}}
D^{\Lambda_{1}}_{m_{\Lambda_{1}}m_{\Lambda_{1}^{\prime}}
^{\prime}}(\sigma)
\sum_{b,b^{\prime}\in B_{N}(m,n)}
D^{\gamma}_{A,m_{A}j, m_{A}^{\prime}i}(b^{\ast})
D^{\gamma^{\prime}}_{A^{\prime},m_{A^{\prime}} 
j^{\prime},m_{A^{\prime}}^{\prime}i^{\prime}}(b^{\prime \ast})
tr_{m,n}(b^{\dagger}\sigma b^{\prime}\sigma^{-1}), 
\hspace{0.4cm}
\label{calculationofcorrelator}
\end{eqnarray}
where we have used the fact that 
the Clebsch-Gordan coefficient 
$C^{\tau_{\Lambda_{1}},\Lambda_{1},m_{\Lambda_{1}}}
_{A,m_{A},A,m_{A}^{\prime}}$ 
and the representation matrix of Brauer elements
are real. 

Let us work on $tr_{m,n}(b\sigma b^{\prime}\sigma^{-1})$. 
Using the Schur-Weyl duality (\ref{swbrauer}), 
we find
\begin{eqnarray}
tr_{m,n}(b^{\dagger}\sigma b^{\prime}\sigma^{-1})
&=&\sum_{\gamma^{\prime\prime}}
t^{\gamma^{\prime\prime}}
\chi^{\gamma^{\prime\prime}}
(b^{\dagger}\sigma b^{\prime}\sigma^{-1}) ,
\end{eqnarray}
where $t^{\gamma^{\prime\prime}}$ and $\chi^{\gamma^{\prime\prime}}(b)$
are the dimension of $U(N)$ and 
the character of the Brauer algebra 
associated 
with the irreducible representation $\gamma^{\prime\prime}$. 
We decompose the character by inserting 
the completeness of 
$|\gamma \rightarrow A,m_{A},i_{A} \rangle $
as
\begin{eqnarray}
\chi^{\gamma^{\prime\prime}}
(b^{\dagger}\sigma b^{\prime}\sigma^{-1}) 
&=&
\langle \gamma^{\prime\prime} \rightarrow B,m_{B},i_{B} |b^{\dagger}
|\gamma^{\prime\prime} \rightarrow C,m_{C},i_{C} \rangle 
D_{m_{C}m_{C}^{\prime}}^{C}(\sigma)
\cr
&&
\langle \gamma^{\prime\prime} \rightarrow C,m_{C}^{\prime},i_{C} |b^{\prime}
|\gamma^{\prime\prime} \rightarrow B,m_{B}^{\prime},i_{B} \rangle 
D_{m_{B}^{\prime}m_{B}}^{B}(\sigma^{-1}), 
\end{eqnarray}
where the summation is assumed to be taken for all repeated letters 
except $\gamma^{\prime\prime}$. 
We note that 
\begin{eqnarray}
\langle \gamma^{\prime\prime} \rightarrow B,m_{B},i_{B} |\sigma
|\gamma^{\prime\prime} \rightarrow C,m_{C},i_{C} \rangle 
=D_{m_{B}m_{C}}^{B}(\sigma)
\delta_{BC}\delta_{i_{B}i_{C}}. 
\end{eqnarray}
Using the above decomposition of the character, 
one can calculate 
(\ref{calculationofcorrelator}) as 
\begin{eqnarray}
&&\langle 
{\cal O}^{\gamma,A,ij,\tau_{\Lambda_{1}}}
_{\Lambda,M_{\Lambda},\Lambda_{1},\tau}{}^{\dagger}
{\cal O}^{\gamma^{\prime},A^{\prime},i^{\prime}j^{\prime},
\tau_{\Lambda_{1}^{\prime}}^{\prime}}
_{\Lambda^{\prime},M_{\Lambda^{\prime}}^{\prime},
\Lambda_{1}^{\prime},\tau^{\prime}}
\rangle \cr
&=&
\delta_{\Lambda \Lambda^{\prime}}
\delta_{M_{\Lambda}M_{\Lambda^{\prime}}^{\prime}}
\delta_{\Lambda_{1}\Lambda_{1}^{\prime}}
\delta_{\tau\tau^{\prime}} 
C^{\tau_{\Lambda_{1}},\Lambda_{1},m_{\Lambda_{1}}}
_{A,m_{A},A,m_{A}^{\prime}}
C^{\tau_{\Lambda_{1}^{\prime}}^{\prime},\Lambda_{1}^{\prime},
m_{\Lambda_{1}^{\prime}}^{\prime}}
_{A^{\prime},m_{A^{\prime}},A^{\prime},m_{A^{\prime}}^{\prime}}
\cr
&&
\sum_{\sigma \in S_{m}\times S_{n}}
D^{\Lambda_{1}}_{m_{\Lambda_{1}}m_{\Lambda_{1}^{\prime}}
^{\prime}}(\sigma)
\sum_{b,b^{\prime}}
D^{\gamma}_{A,m_{A}j, m_{A}^{\prime}i}(b^{\ast})
D^{\gamma^{\prime}}_{A^{\prime},m_{A^{\prime}} j^{\prime},
m_{A^{\prime}}^{\prime}i^{\prime}}(b^{\prime \ast})
\cr
&&
t^{\gamma^{\prime\prime}}
\langle \gamma^{\prime\prime} \rightarrow B,m_{B},i_{B} |b^{\dagger}
|\gamma^{\prime\prime} \rightarrow C,m_{C},i_{C} \rangle 
D_{m_{C}m_{C}^{\prime}}^{C}(\sigma)
 \cr
&&
\langle \gamma^{\prime\prime} \rightarrow C,m_{C}^{\prime},i_{C} |b^{\prime}
|\gamma^{\prime\prime} \rightarrow B,m_{B}^{\prime},i_{B} \rangle 
D_{m_{B}^{\prime}m_{B}}^{B}(\sigma^{-1})
\cr
&=&
\delta_{\Lambda \Lambda^{\prime}}
\delta_{M_{\Lambda}M_{\Lambda^{\prime}}^{\prime}}
\delta_{\Lambda_{1}\Lambda_{1}^{\prime}}
\delta_{\tau\tau^{\prime}} 
C^{\tau_{\Lambda_{1}},\Lambda_{1},m_{\Lambda_{1}}}
_{A,m_{A},A,m_{A}^{\prime}}
C^{\tau_{\Lambda_{1}^{\prime}}^{\prime},
\Lambda_{1}^{\prime},m_{\Lambda_{1}^{\prime}}^{\prime}}
_{A^{\prime},m_{A^{\prime}},A^{\prime},m_{A^{\prime}}^{\prime}}
\cr
&&
\frac{1}{t^{\gamma}}\delta_{\gamma\gamma^{\prime}}
\delta_{AA^{\prime}}
\sum_{\sigma \in S_{m}\times S_{n}}
D^{\Lambda_{1}}_{m_{\Lambda_{1}}m_{\Lambda_{1}^{\prime}}
^{\prime}}(\sigma)
D_{m_{A}^{\prime}m_{A^{\prime}}^{\prime}}^{A}(\sigma)
D_{m_{A^{\prime}}m_{A}}^{A}(\sigma^{-1})
\delta_{i^{\prime}i}\delta_{jj^{\prime}}
\cr
&=&
\delta_{\Lambda \Lambda^{\prime}}
\delta_{M_{\Lambda}M_{\Lambda^{\prime}}^{\prime}}
\delta_{\Lambda_{1}\Lambda_{1}^{\prime}}
\delta_{\tau\tau^{\prime}} 
\delta_{AA^{\prime}}
C^{\tau_{\Lambda_{1}},\Lambda_{1},m_{\Lambda_{1}}}
_{A,m_{A},A,m_{A}^{\prime}}
C^{\tau_{\Lambda_{1}^{\prime}}^{\prime},
\Lambda_{1}^{\prime},m_{\Lambda_{1}^{\prime}}^{\prime}}
_{A^{\prime},m_{A^{\prime}},A^{\prime},m_{A^{\prime}}^{\prime}}
\cr
&&
\frac{1}{t^{\gamma}}\delta_{\gamma\gamma^{\prime}}
\frac{m!n!}{d_{\Lambda_{1}}}
\sum_{\tau_{\Lambda_{1}}^{\prime}}
\delta_{ii^{\prime}}\delta_{jj^{\prime}}
C^{\tau_{\Lambda_{1}}^{\prime},\Lambda_{1},m_{\Lambda_{1}}}
_{A,m_{A}^{\prime},A,m_{A}}
C^{\tau_{\Lambda_{1}}^{\prime},\Lambda_{1},
m_{\Lambda_{1}^{\prime}}^{\prime}}
_{A,m_{A^{\prime}}^{\prime},A,m_{A^{\prime}}}
\cr
&=&
\delta_{\Lambda \Lambda^{\prime}}
\delta_{M_{\Lambda}M_{\Lambda^{\prime}}^{\prime}}
\delta_{\Lambda_{1}\Lambda_{1}^{\prime}}
\delta_{\tau\tau^{\prime}} 
\delta_{AA^{\prime}}
\delta_{
\tau_{\Lambda_{1}}
\tau_{\Lambda_{1}^{\prime}}^{\prime}}
\delta_{ii^{\prime}}\delta_{jj^{\prime}}
m!n!d_{\Lambda_{1}}
\frac{1}{t^{\gamma}}\delta_{\gamma\gamma^{\prime}}. 
\label{calculationoftwopoint}
\end{eqnarray}
The second equality 
was obtained by performing 
the summation for $b$ and $b^{\prime}$
using the formula 
derived in subsection \ref{sec:proofofDD}.
To show the third equality, we have used 
\begin{eqnarray}
&&
\sum_{\sigma \in S_{m}\times S_{n}}
D^{\Lambda_{1}}_{m_{\Lambda_{1}}m_{\Lambda_{1}^{\prime}}
^{\prime}}(\sigma)
D_{m_{A}^{\prime}m_{A^{\prime}}^{\prime}}^{A}(\sigma)
D_{m_{A^{\prime}}m_{A}}^{A}(\sigma^{-1})
\cr
&=&
\sum_{\sigma \in S_{m}\times S_{n}}
D^{\Lambda_{1}}_{m_{\Lambda_{1}}
m_{\Lambda_{1}^{\prime}}^{\prime}}(\sigma)
D_{m_{A}^{\prime}m_{A^{\prime}}^{\prime}}^{A}(\sigma)
D_{m_{A}m_{A^{\prime}}}^{A}(\sigma)
\cr
&=&
\frac{m!n!}{d_{\Lambda_{1}}}
\sum_{\tau_{\Lambda_{1}}^{\prime}}
C^{\tau_{\Lambda_{1}}^{\prime},\Lambda_{1},m_{\Lambda_{1}}}
_{A,m_{A}^{\prime},A,m_{A}}
C^{\tau_{\Lambda_{1}}^{\prime},\Lambda_{1},
m_{\Lambda_{1}^{\prime}}^{\prime}}
_{A,m_{A^{\prime}}^{\prime},A,m_{A^{\prime}}}. 
\end{eqnarray}
This is a generalisation of (163) in \cite{0711.0176}. 
The forth equality of (\ref{calculationoftwopoint})
follows from the orthogonality of Clebsch-Gordan coefficient 
\cite{Hamermesh}:
\begin{eqnarray}
\sum_{m_{A},m_{A}^{\prime}}C^{\tau_{\Lambda_{1}},\Lambda_{1},m_{\Lambda_{1}}}
_{A,m_{A},A,m_{A}^{\prime}}
C^{\tau_{\Lambda_{1}^{\prime}},\Lambda_{1}^{\prime},m_{\Lambda_{1}^{\prime}}}
_{A,m_{A},A,m_{A}^{\prime}}
=\delta_{\tau_{\Lambda_{1}},\tau_{\Lambda_{1}^{\prime}}}
\delta_{\Lambda_{1},\Lambda_{1}^{\prime}}
\delta_{m_{\Lambda_{1}},m_{\Lambda_{1}^{\prime}}}. 
\end{eqnarray}

\subsection{A formula}
\label{sec:proofofDD}
We now calculate
\begin{eqnarray}
&&\sum_{b\in B_{N}(m,n)}
\langle \gamma \rightarrow A,m_{A},i_{A} |
b^{\ast}
|\gamma \rightarrow B,m_{B},i_{B} \rangle 
\langle \gamma^{\prime} \rightarrow C,m_{C},i_{C} |
b
|\gamma^{\prime} \rightarrow D,m_{D},i_{D} \rangle .
\hspace{0.6cm}
\end{eqnarray}
This can be worked out as 
\begin{eqnarray}
&&
\sum_{b\in B_{N}(m,n)}\sum_{I,J,K,L}
B_{\gamma I;A,m_{A},i_{A}}^{\dagger}
B_{\gamma J;B,m_{B},i_{B}}
D^{\gamma}_{IJ}(b^{\ast})
B_{\gamma^{\prime} K;C,m_{C},i_{C}}^{\dagger}
B_{\gamma^{\prime} L;D,m_{D},i_{D}}
D^{\gamma^{\prime}}_{KL}(b) \cr
&=&
\frac{1}{t^{\gamma}}\delta_{\gamma\gamma^{\prime}}
\sum_{I,J}
B_{\gamma I;A,m_{A},i_{A}}^{\dagger}
B_{\gamma J;B,m_{B},i_{B}}
B_{\gamma J;C,m_{C},i_{C}}^{\dagger}
B_{\gamma I;D,m_{D},i_{D}} \cr
&=&
\frac{1}{t^{\gamma}}
\delta_{\gamma\gamma^{\prime}}
\delta_{AD}\delta_{BC}
\delta_{m_{A}m_{D}}\delta_{m_{B}m_{C}}
\delta_{i_{A}i_{D}}\delta_{i_{B}i_{C}}. 
\end{eqnarray}
Some equations needed to show this 
are collected in the following. 
\begin{itemize}
\item 
A representation matrix for an element $b$ in the Brauer 
algebra can be denoted by 
$D^{\gamma}_{IJ}(b)=
\langle \gamma ,I |b|\gamma ,j \rangle$ 
($1\le I,J\le d_{\gamma}$). 
$d_{\gamma}$ is the dimension of the Brauer algebra 
relevant to  
the irreducible representation $\gamma$. 
The basis is complete:
\begin{eqnarray}
\sum_{I=1}^{d_{\gamma}}|\gamma ,I \rangle \langle \gamma ,I |=1. 
\end{eqnarray}

The orthogonality of the representation matrix is \cite{ramthesis}
\begin{eqnarray}
\sum_{b\in B_{N}(m,n)}
D^{\gamma}_{IJ}(b^{\ast})D^{\gamma^{\prime}}_{KL}(b)
=\frac{1}{t^{\gamma}}\delta_{JK}\delta_{IL}\delta^{\gamma\gamma^{\prime}}.
\label{orthogonalrepresentationbrauer}
\end{eqnarray}

\item 
We introduce 
the branching coefficient $B_{\gamma I;A,m_{A},i}$ as 
\begin{eqnarray}
B_{\gamma I;A,m_{A},i}=\langle \gamma,I| 
\gamma \rightarrow A,m_{A},i\rangle. 
\label{defBranchingcoefficient}
\end{eqnarray}
The orthogonality of the branching coefficient can be derived as 
\begin{eqnarray}
\sum_{I}  (  B_{\gamma  , I; A , m_A  ,i_A })^{\dagger}
B_{\gamma , I ; B , m_B , i_B}
&=&
\sum_{I}
\langle \gamma \rightarrow A,m_{A},i_A  | \gamma, I \rangle 
\langle \gamma, I |\gamma \rightarrow  B ,m_{B},i_B \rangle \cr
&=& \delta_{AB } \delta_{m_A m_B } \delta_{i_A i_B }, 
\end{eqnarray}
where we have chosen an orthogonal basis 
\begin{eqnarray}
\langle \gamma \rightarrow B ,m_{B}, k
|\gamma \rightarrow A ,m_{A}, j\rangle 
=\delta_{AB}\delta_{m_{A}m_{B}}\delta_{jk}. 
\end{eqnarray}

\end{itemize}


\section{Two actions of the symmetric group }
\setcounter{equation}{0} 
\label{sec:symmetryCG}
We introduced two Brauer algebras to label the orthogonal basis. 
One was responsible for flavour indices and the other was for 
colour indices. 
They have the same sub-algebra, i.e. the group algebra of 
the symmetric group $S_{m}\times S_{n}$, but they 
allow the sub-algebra to act on the operator
in different ways. 
The two different ways 
are correlated as we shall see in 
(\ref{relationtwosymmetric}). 
Because 
the orthogonal basis 
should respect it, 
the Clebsh-Gordan coefficient should also satisfy 
a relation. 
We shall see that this is the case. 

The symmetric group associated with colour indices 
act on the covariant operator as 
\begin{eqnarray}
\sigma \left
(O_{\Lambda,M_{\Lambda},\Lambda_{1},m_{\Lambda_{1}},\tau}
\right)^{I}_{J}
\sigma^{-1}
&=&\left(O_{\Lambda,M_{\Lambda},\Lambda_{1},m_{\Lambda_{1}},\tau}
\right)^{\sigma(I)}_{\sigma(J)}
\cr
&=&
\sum_{\vec{a},\vec{b}}C_{\Lambda,M_{\Lambda},
\Lambda_{1},m_{\Lambda_{1}},\tau}^{\vec{a},\vec{b}}
\left(O_{\vec{a},\vec{b}}
\right)^{\sigma(I)}_{\sigma(J)}. 
\end{eqnarray}
On the other hand, 
the other symmetric group associated with flavour indices 
act on the covariant operator as 
\begin{eqnarray}
D_{m_{\Lambda_{1}}m_{\Lambda_{1}}^{\prime}}^{\Lambda_{1}}(\sigma)
O_{\Lambda,M_{\Lambda},\Lambda_{1},m_{\Lambda_{1}}^{\prime},\tau}
&=&
\sum_{\vec{a},\vec{b}}C_{\Lambda,M_{\Lambda},
\Lambda_{1},m_{\Lambda_{1}},\tau}^{\sigma(\vec{a}),\sigma(\vec{b})}
\left(O_{\vec{a},\vec{b}}
\right) \cr
&=&
\sum_{\vec{a},\vec{b}}C_{\Lambda,M_{\Lambda},
\Lambda_{1},m_{\Lambda_{1}},\tau}^{\vec{a},\vec{b}}
\left(O_{\sigma^{-1}(\vec{a}),\sigma^{-1}(\vec{b})}
\right) .
\end{eqnarray}
These two actions are not independent because of 
\begin{eqnarray}
\left(O_{\vec{a},\vec{b}}
\right)^{\sigma(I)}_{\sigma(J)}=
\left(O_{\sigma(\vec{a}),\sigma(\vec{b})}
\right)^{I}_{J}. 
\label{relationtwosymmetric}
\end{eqnarray}
This means the following equation to the covariant operator
\begin{eqnarray}
O_{\Lambda,M_{\Lambda},\Lambda_{1},m_{\Lambda_{1}},\tau}
=D_{m_{\Lambda_{1}}m_{\Lambda_{1}}^{\prime}}^{\Lambda_{1}}(\sigma)
\sigma \left
(O_{\Lambda,M_{\Lambda},\Lambda_{1},m_{\Lambda_{1}}^{\prime},\tau}
\right)
\sigma^{-1}. 
\label{symcovariant}
\end{eqnarray}
Substituting this in 
the gauge invariant operator (\ref{gaugeinvariant}), 
we obtain
\begin{eqnarray*}
{\cal O}_{\Lambda,M_{\Lambda},\Lambda_{1},\tau}
^{\gamma,A,ij,\tau_{\Lambda_{1}}}
&=&
C^{\tau_{\Lambda_{1}},\Lambda_{1},m_{\Lambda_{1}}}_{A,m_{A},A,m_{A}^{\prime}}
\sum_{b}
D^{\gamma}_{A,m_{A}i,m_{A}^{\prime}j}(b^{\ast})
D_{m_{\Lambda_{1}}m_{\Lambda_{1}}^{\prime}}^{\Lambda_{1}}(\sigma)
tr_{m,n}(b\sigma 
O_{\Lambda,M_{\Lambda},\Lambda_{1},m_{\Lambda_{1}}^{\prime},\tau}
\sigma^{-1})
\cr
&=&
C^{\tau_{\Lambda_{1}},\Lambda_{1},m_{\Lambda_{1}}}_{A,m_{A},A,m_{A}^{\prime}}
\sum_{b}
D^{\gamma}_{A,m_{A}i,m_{A}^{\prime}j}((\sigma b \sigma^{-1})^{\ast})
D_{m_{\Lambda_{1}}m_{\Lambda_{1}}^{\prime}}^{\Lambda_{1}}(\sigma)
tr_{m,n}(b
O_{\Lambda,M_{\Lambda},\Lambda_{1},m_{\Lambda_{1}}^{\prime},\tau}).
\end{eqnarray*}
This $\sigma$-dependence can be shown to vanish 
as expected from the consistency of this operator. 
Using 
\footnote{
This comes from 
the expression of $b^{\ast}$ 
in (3.27) of  \cite{0709.2158}.  
with the assist of 
$\Sigma(\sigma b\sigma^{-1})=\sigma \Sigma(b) \sigma^{-1}$, where 
$\Sigma(\sigma b \tau)=\sigma \Sigma(b) \tau$ 
is not true. 
}
\begin{eqnarray}
(\sigma b \sigma^{-1})^{\ast}=\sigma b^{\ast} \sigma^{-1}, 
\end{eqnarray}
we can show that the $\sigma$-dependence disappears as 
\begin{eqnarray}
&&
C^{\tau_{\Lambda_{1}},\Lambda_{1},m_{\Lambda_{1}}}_{A,m_{A},A,m_{A}^{\prime}}
\sum_{b}
D^{\gamma}_{A,m_{A}i,m_{A}^{\prime}j}((\sigma b \sigma^{-1})^{\ast})
D_{m_{\Lambda_{1}}m_{\Lambda_{1}}^{\prime}}^{\Lambda_{1}}(\sigma)
\cr
&=&
C^{\tau_{\Lambda_{1}},\Lambda_{1},m_{\Lambda_{1}}}_{A,m_{A},A,m_{A}^{\prime}}
\sum_{b}
D^{A}_{m_{A}m_{A}^{\prime\prime}}(\sigma)
D^{\gamma}_{A,m_{A}^{\prime\prime}i,m_{A}^{\prime\prime\prime}j}(b ^{\ast})
D^{A}_{m_{A}^{\prime\prime\prime}m_{A}^{\prime}}(\sigma^{-1})
D_{m_{\Lambda_{1}}m_{\Lambda_{1}}^{\prime}}^{\Lambda_{1}}(\sigma)
\cr
&=&
C^{\tau_{\Lambda_{1}},\Lambda_{1},m_{\Lambda_{1}}^{\prime\prime}}
_{A,m_{A}^{\prime\prime},A,m_{A}^{\prime\prime\prime}}
D^{\Lambda_{1}}_{m_{\Lambda_{1}}^{\prime\prime}m_{\Lambda_{1}}}(\sigma^{-1})
D_{m_{\Lambda_{1}}m_{\Lambda_{1}}^{\prime}}^{\Lambda_{1}}(\sigma)
\sum_{b}D^{\gamma}_{A,m_{A}^{\prime\prime}i,m_{A}^{\prime\prime\prime}j}(b ^{\ast})
\cr
&=&
C^{\tau_{\Lambda_{1}},\Lambda_{1},m_{\Lambda_{1}}^{\prime}}
_{A,m_{A}^{\prime\prime},A,m_{A}^{\prime\prime\prime}}
\sum_{b}D^{\gamma}_{A,m_{A}^{\prime\prime}i,m_{A}^{\prime\prime\prime}j}(b ^{\ast}). 
\end{eqnarray}
The final step comes from the following equation 
\begin{eqnarray}
&&
C^{\tau_{\Lambda_{1}},\Lambda_{1},m_{\Lambda_{1}}}_{A,m_{A},A,m_{A}^{\prime}}
D^{A}_{m_{A}m_{A}^{\prime\prime}}(\sigma)
D^{A}_{m_{A}^{\prime\prime\prime}m_{A}^{\prime}}(\sigma^{-1})
\cr
&=&
C^{\tau_{\Lambda_{1}},\Lambda_{1},m_{\Lambda_{1}}}_{A,m_{A},A,m_{A}^{\prime}}
D^{A}_{m_{A}^{\prime\prime}m_{A}}(\sigma^{-1})
D^{A}_{m_{A}^{\prime\prime\prime}m_{A}^{\prime}}(\sigma^{-1})
\cr
&=&
C^{\tau_{\Lambda_{1}},\Lambda_{1},m_{\Lambda_{1}}^{\prime\prime}}
_{A,m_{A}^{\prime\prime},A,m_{A}^{\prime\prime\prime}}
D^{\Lambda_{1}}_{m_{\Lambda_{1}}^{\prime\prime}m_{\Lambda_{1}}}(\sigma^{-1}).
\end{eqnarray}
See appendix A in \cite{0711.0176} for the derivation.



\end{document}